\documentclass[a4paper,11pt]{article}
\pdfoutput=1 

\usepackage{jheppub} 

\usepackage[T1]{fontenc} 
\usepackage{caption}
\usepackage{comment}
\usepackage{amsmath}
\usepackage{graphicx}
\usepackage{tikz}
\usetikzlibrary{arrows,decorations.pathmorphing}
\usepackage[compat=1.1.0]{tikz-feynman}
\usepackage{amssymb}
\usepackage{bm,commath,mathtools,wasysym,xfrac}
\usepackage{pdfpages}
\usepackage{tcolorbox}
\usepackage{braket}
\usepackage{MnSymbol}
\usepackage[english]{babel} 
\usepackage{blindtext}

\newcommand{\bq}{\bar{q}}

\usepackage{cancel}
\usepackage{color}
\usepackage{float}

\title{\boldmath On the classical limit of the (sub)$^{n}$-leading soft graviton theorems in $D = 4$ without deflection}

\author[a,b]{Samim Akhtar}

\affiliation[a]{The Institute of Mathematical Sciences, \\
	IV Cross Road, C.I.T. Campus, Taramani, Chennai 600 113, India}
\affiliation[b]{Homi Bhabha National Institute, \\
	Training School Complex, Anushakti Nagar, Mumbai 400 094, India}

\emailAdd{samimakhtar@imsc.res.in}

\abstract{Tree-level gravitational amplitudes satisfy an infinite hierarchy of soft factorization theorems. The existence of these theorems has been recently linked with the existence of an infinite tower of asymptotic symmetries. In this paper, we analyze the relevance of the soft graviton theorems beyond sub-leading order in the context of classical gravitational scattering in four dimensions. More in detail, we show that the infinite impact parameter limit of the late-time gravitational field emitted during a classical scattering can be derived using these factorization theorems. The classical field obtained in this (infinite impact parameter) regime has an expansion in the frequency of the detector where the modes scale as $\omega^{n}\log{\omega}$ with a vanishing memory.}

\makeatletter
\gdef\@fpheader{}
\makeatother

\begin{document} 
	\maketitle
	\flushbottom
	
	\section{Introduction}
The two most prominent perturbative techniques to analyze classical gravitational scattering of two objects whose Schwarzschild radii are parametrically small compared to the impact parameter of the scattering are the post-Minkowskian (PM) and post-Newtonian (PN) schemes \cite{Bern:2019nnu,Bern:2020gjj,Bern:2020uwk,Bern:2021dqo,Bern:2021yeh,Bern:2021xze,Bern:2022kto,Bern:2023ccb,Alaverdian:2024spu,Kalin:2019rwq,Cho:2021arx,Kalin:2019inp,Kalin:2020fhe,Kalin:2020lmz,Dlapa:2021npj,Dlapa:2021vgp,Dlapa:2022lmu,Dlapa:2023hsl,Kalin:2020mvi,Liu:2021zxr,Kalin:2022hph,Manohar:2022dea,Cristofoli:2020uzm,DiVecchia:2021ndb,DiVecchia:2021bdo,Alessio:2022kwv,DiVecchia:2022piu,DiVecchia:2022owy,Georgoudis:2023eke,Bjerrum-Bohr:2018xdl,Cristofoli:2019neg,Bjerrum-Bohr:2021vuf,Bjerrum-Bohr:2021din,Bjerrum-Bohr:2021wwt,Dlapa:2024cje,Amalberti:2024jaa,Porto:2024cwd,Bini:2022enm,Bini:2024ijq,Bini:2024tft,Bini:2022wrq,Bini:2021gat,Bini:2020rzn,Bini:2020hmy,Bini:2020nsb,Bini:2020wpo,Bini:2019nra,Bini:2020flp,Bini:2024rsy}. In the former, one computes relevant classical observables such as scattering angle or the gravitational waveform as a perturbation series in the gravitational constant $G$. On the other hand, PN expansion can be used when the incoming velocities of the scattering particles or objects in a bound orbit are non-relativistic. The two perturbative schemes are intricately tied to each other in the case of large impact parameter as beautifully explained in \cite{Bern:2019crd,Bern:2020buy} and are the most potent tools in analyzing the relationship between quantum amplitudes and classical scattering. However, a complementary perturbative expansion leads to different insights about gravitational radiation emitted in a scattering process. This expansion is not in terms of parameters intrinsic to the scattering process but the characteristic frequency of the detector placed at null infinity. It is known as the soft expansion of gravitational radiation. At any given order in soft expansion, the radiative field is exact to all orders in PM and PN expansion. It is hence a non-perturbative probe to gravitational scattering and offers remarkable insights into universal modes of gravitational radiation in classical scattering \cite{Sen:2017nim,Laddha:2017ygw,Laddha:2018rle,Laddha:2018myi,Laddha:2018vbn,Sahoo:2018lxl,Saha:2019tub,Laddha:2019yaj,Sahoo:2020ryf,Sahoo:2021ctw,Ghosh:2021bam,Krishna:2023fxg,Sen:2024bax,Chakrabarti:2017ltl,AtulBhatkar:2018kfi}.

It is  by now well known that for a generic gravitational scattering, the radiative field has the following form under soft expansion,
\begin{align} \label{softexp}
h_{\mu\nu}(\omega, r, \hat{n})\, &=\, \frac{1}{r} e^{i\omega r}\, \sum_{N=-1}^{\infty}\, \omega^{N}\, h_{\mu\nu}^{N}(\hat{n})\, +\, \sum_{m=0}^{\infty}\, \omega^{m}\, (\log{\omega})^{m+1}\, h_{\mu\nu}^{\log \, m}(\hat{n})\, \cr
&\hspace{3cm} +\, \sum_{N, M\vert\, M - N > -1} 
\omega^{M} (\log{\omega})^{N}\, h_{\mu\nu}^{\log (N, M)}(\hat{n}) + \mathcal{O}\left(\frac{1}{r^{2}}\right),
\end{align}
where $\hat{n}$ is the unit vector on the celestial sphere and all the terms subleading in $\frac{1}{r}$ do not contribute to the radiative flux at null infinity \cite{Geiller:2022vto,Geiller:2024amx,Geiller:2024ryw}. 

The classification of soft terms into three distinct families in equation \ref{softexp} is as follows. The first one is a Laurent expansion in the detector frequency. The second set of terms that scale as $\omega^{N}\ln\omega^{N+1}$ capture the (infinite) set of leading logarithmic terms at all orders in the soft expansion and were discovered in a series of papers \cite{Laddha:2018myi,Sahoo:2018lxl,Saha:2019tub,Sahoo:2020ryf,Sahoo:2021ctw,Ghosh:2021bam}. Finally, the third family encapsulates the logarithmic terms which (for a given $N$) are sub-leading relative to the corresponding leading logarithmic soft factors.

The reason for such a classification is intricately tied to the idea of universality inherent in soft expansion. Given a set of incoming and outgoing momenta and spins of the scattering objects, if a specific mode in the soft expansion does not depend on the details of the underlying equations of motion and the details of the scattering, then we refer to such a mode as a universal term in the soft expansion. It has been known since the early 60s that in a generic classical gravitational scattering in which massive objects with arbitrary multipole moments emit gravitational radiation, then the leading soft factor $h_{\mu\nu}^{-1}(\hat{n})$ is universal and only depends on the incoming and outgoing momenta of the scattering particles\footnote{We here assume that massless particles model finite energy gravitational radiation that can also emit soft radiation. In the literature, the contribution of massive states to the leading soft factor is known as linear memory and the contribution of massless particles, including gravitons, to the leading soft factor is known as non-linear or null memory \cite{Zeldovich:1974gvh,Christodoulou:1991cr,Wiseman:1991ss,Thorne:1992sdb,Bieri:2013ada,Bieri:2015yia}.}. In last ten years, B. Sahoo, A. Sen, and their collaborators have shown that universality of gravitational radiation in the soft expansion is not only tied to the leading soft factor. It has been now rigorously shown that the leading logs $N \leq\, 2$ are universal and only depend on the momenta of the incoming and outgoing massive objects \cite{Sen:2017nim,Laddha:2018myi,Sahoo:2018lxl,Saha:2019tub,Sahoo:2020ryf,Sahoo:2021ctw,Ghosh:2021bam,Alessio:2024onn}. It has been conjectured that the universality extends to all $N$ and a specific formula for the coefficient of $\omega^{N}\ln\omega^{N+1}$ has been put forward by Heissenberg et al. in the case of $2-2$ scattering \cite{Alessio:2024onn}. 

An interesting aspect of the leading log soft factor $h_{\mu\nu}^{\log}(\hat{n})$ is that, unlike $h_{\mu\nu}^{-1}$, it is sourced even in the limit when $n$ particles which undergo scattering are mutually so far apart that each of them experiences a vanishing deflection! The source of such a radiative mode then is the asymptotic interaction between the incoming or outgoing states, leading to the emission of gravitational radiation only from $t\, \rightarrow\, \pm \infty$. In this paper, we analyze the classical gravitational radiation in $D = 4$ dimensions where two massive objects with momenta $p_{1}, p_{2}$ scatter at finite impact parameter. We then analyze the soft expansion of gravitational radiation in $\vert b\vert \rightarrow\, \infty$ limit such that $\omega \vert b\vert$ is fixed. As we show below, in this regime, all the terms of the form $\omega^{m}\log{\omega} \vert m \geq\, 1$ survive and can be completely determined by the so-called (sub)$^{n}$-leading soft graviton theorems for tree-level gravitational amplitudes. These theorems reveal the extent to which a gravitational amplitude factorizes when one of the gravitons becomes soft as compared to other external momenta. This is discussed in more detail in section \ref{sec:treesoft}. For any $n \geq 1$, (sub)$^{n}$-leading soft expansion of the tree-level scattering amplitude for a generic theory of gravity with arbitrary matter coupling has the following form:
 \begin{align}
\lim_{\omega\, \rightarrow\, 0} \partial_{\omega}^n \omega{\cal A}_{5}(\tilde{p}_{1},\tilde{p}_{2} \rightarrow\, p_{1},p_{2},k)\, =\, \hat{S}^{n} {\cal A}_{4} + {\cal B}_{n}(p_{1}, p_{2}, p_{1}^{\prime}, p_{2}^{\prime}, \hat{k})\, {\cal A}_{4} + {\cal R}_{n}(p_{1}, p_{2}, p_{1}^{\prime}, p_{2}^{\prime}, \hat{k}) \,,
 \end{align}
where ${\cal B}_{n}$ is the non-universal part of the factorization formula that depends on the irrelevant three-point couplings in the theory \cite{Elvang:2016qvq,Laddha:2017ygw} and ${\cal R}^{n}\, \neq \, 0\, \forall\, n > 2$. Hence the soft expansion of the tree-level amplitude does not factorize. However, the tensorial structure of the remainder term ${\cal R}_{n}$, which is linear in the polarization of the soft graviton leads to the following ``factorization formula'' for tree-level amplitude at all orders in the soft expansion \cite{Campiglia:2018dyi}:
\begin{align} 
\lim_{\omega\, \rightarrow\, 0} \partial_{\omega}^n \omega \Pi^{-}_{\hat{n}}{\cal A}_{5}(\tilde{p}_{1},\tilde{p}_{2} \rightarrow\, p_{1},p_{2},k)\, =\, \Pi^{-}_{\hat{n}} \hat{S}^{n} {\cal A}_{4}(\tilde{p}_{1},\tilde{p}_{2} \rightarrow\, p_{1},p_{2})\,,
\end{align} 
where in the present paper we consider scattering of two scalar particles minimally coupled to gravity, hence ${\cal B}_{n} = 0\, \forall \, n$. $\Pi^{-}_{\hat{n}} := D_{z}^{n+1} (1 + |z|^2)^{-1}$ is the projection operator, where $\hat{k} = (1,\hat{n}(z,\bar{z}))$ and $z,\bar{z}$ are the stereographic coordinates. 

We thus see that higher-order soft theorems appear to be rather limited in their ability to constrain the gravitational scattering as they are only a component of the radiative field which is ``orthogonal'' to the remainder term. In this paper, we prove that despite their limitations, the (sub)$^{n}$-leading soft theorems can capture all the logarithmic soft factors in the limit of vanishing deflection.  As these higher-order tree-level soft theorems are intricately tied to the discovery of the $w_{1+\infty}$ asymptotic symmetry algebra \cite{Freidel:2021ytz,Guevara:2021abz,Strominger:2021mtt,Geiller:2024bgf}, our analysis can potentially reveal the link between higher spin asymptotic symmetries and a subset of logarithmic terms in the soft expansion of gravitational radiation.

This paper is organized as follows. In section \ref{kmocsection}, we review the KMOC formalism used to compute classical observables from scattering amplitudes. Next, we briefly review the classical log soft theorems in section \ref{sec:classicallogs}. Then we review the infinite hierarchy of soft graviton theorems for tree-level amplitudes in section \ref{sec:treesoft}. In section \ref{sgravitysec}, we compute the radiative field for the scattering of scalar particles via gravitational interaction using the KMOC formalism and take the soft expansion. In section \ref{subnsection}, we compare the soft radiation obtained via the quantum soft theorems with the soft expansion of the radiation kernel to (sub)$^n$-leading order, extract out the logarithmic contributions, and identify the remainder terms. We then summarise the important results for $n=3$ i.e. the (sub)$^3$-leading order in the soft expansion in section \ref{sub3section}. We conclude by discussing some open questions in section \ref{discussion}. In the appendices we state our conventions, review the calculation of obtaining the logarithmic contributions to the soft radiation at (sub)$^2$-leading order in frequency, and evaluate some integrals that are used in the main text.
\section{KMOC formalism in a nutshell} \label{kmocsection}
	The KMOC formalism \cite{Kosower:2018adc,Cristofoli:2021vyo} is a framework used to compute classical observables, such as linear impulse or radiation emitted from on-shell scattering amplitudes for large impact parameter scattering\footnote{In this scattering configuration, the particles don't deviate significantly from the initial trajectories, with the impact parameter determining the characteristic length scale.}. The basic idea is to take
wave packets for incoming (classical) particles, evolve them using a quantum S-matrix operator, and then compute the expectation value of an observable in the final state. An appropriate classical limit is then applied to obtain the classical result. The primary characteristic of the formalism is that the classical limit is taken before evaluating the full amplitude which makes the computation substantially simpler. Additionally, radiation reaction effects are inherently encoded within the framework. For a short sample of the results obtained with the formalism, we refer the reader to \cite{Bautista:2021llr,Elkhidir:2023dco,Bern:2021xze,Akhtar:2024mbg,A:2022wsk}.
	
	The initial state is described as:  
	\begin{equation}\label{ini}
		\ket{\Psi} =\int\prod_{i=1}^{2}d\Phi(p_{i}) e^{ip_{2}\cdot b/\hbar}\phi_{i}(p_{i}) \ket{\vec{p}_{1};\vec{p}_{2}},
		\end{equation}
 where
	\begin{equation}
		d\Phi(p) = \frac{d^{4}p}{(2\pi)^{4}}\hat{\delta}(p^{2}-m^{2})\Theta(p^{0}), \  \int d\Phi(p)\ |\phi(p)|^{2} = 1,
	\end{equation}
	$\phi_{i}(p_{i})$s are the minimum uncertainty wave packets for the particles. The wavepacket of the second particle is translated, with respect to the first particle's wavepacket, by a distance of $b$ - the impact parameter. As the initial particles are described by coherent states we have
\begin{equation}
\braket{\mathbb{P}^{\mu}_{i}} = m_{i}u^{\mu}_{i}+\mathcal{O}(\hbar),\qquad \frac{\sigma_{i}^{2}}{m_{i}^{2}} = \frac{(\braket{\mathbb{P}^{2}_{i}} - \braket{\mathbb{P}_{i}}^{2})}{m_{i}^{2}} \xrightarrow{\hbar \rightarrow 0} 0,
\end{equation}
	where $\sigma_{i}^{2}$ is the variance and $m_{i}$s are the masses of the particles. Here the expectation value of the momentum operator is with respect to the initial state in equation \eqref{ini}.	
For massive spinning particles and more detailed construction of these wavefunctions, we refer the reader to \cite{Aoude:2021oqj}.\\
	The computation of classical observables involves the change in the expectation value of a quantum mechanical operator: 
	\begin{align}\label{3.1}
		\braket{\Delta\mathbb{O}^{A}}=\bra{\Psi}S^{\dag}\mathbb{O}^{A}S\ket{\Psi} - \bra{\Psi}\mathbb{O}^{A}\ket{\Psi}.
	\end{align}
	where $S=I+iT$ is the S-matrix. For the linear impulse, $\mathbb{O}^{A}=\mathbb{P}^{\mu}$, the momentum operator, and $\mathbb{O}^{A}=h^{\mu\nu}(x)$, the graviton field operator, from which we read off the radiation kernel.
	The expression in equation \eqref{3.1} can be simplified using the on-shell completeness relation and unitarity of the S-matrix \cite{Kosower:2018adc}. 
	To compute radiation, an important intermediate quantity introduced by KMOC is the so-called radiation kernel $R^{\mu\nu}
(k)$. Radiation kernel is simply the field radiated at momentum $k^\mu = \omega(1,\hat{n})$ and is a result of in-elastic scattering. The radiation kernel has the following compact expression \cite{Cristofoli:2021vyo}
	\begin{align}
		\langle \mathcal{R}^{\mu\nu}(k) \rangle &=\ \hbar^{3/2}\int \prod_{i=1}^{2} \hat{d}^{4}q_{i}\  \hat{\delta}(2p_{i}\cdot q_{i}+q_{i}^{2})\ e^{-iq_{2}\cdot b/\hbar}\cr
		&\qquad \qquad \hat{\delta}^{(4)}(q_{1}+q_{2}-k)\ \mathcal{A}_{5}^{\mu\nu}(p_{1}+q_{1},p_{2}+q_{2}\rightarrow p_{1},p_{2},k)\  +\ \mathcal{O}(T^{\dag}T)\,,
	\end{align}
where $\mathcal{A}_5^{\mu\nu}$ is the tree-level five-point amplitude.
	The classical limit is taken at the level of the integrand, by expressing massless momenta in terms of their wave numbers (e.g.\ $q_{i}=\bar{q}_{i}\hbar$), appropriately rescaling the dimensionful couplings and keeping leading order terms in $\hbar$. In gravity, the dimensionless coupling is obtained by $\kappa \rightarrow \kappa/\sqrt{\hbar}$, where $\kappa = \sqrt{32\pi G}$. Here $G $ is the gravitational constant. We can now write down the classical limit of the radiation kernel, at leading order in the coupling as follows
	\begin{align} \label{radiationkernel}
		\mathcal{R}^{\mu\nu}(\bar{k})=  \Big \llangle \int \prod_{i=1}^{2} \hat{d}^{4}\bar{q}_{i}\  \hat{\delta}(2p_{i}\cdot \bar{q}_{i}) & \ e^{-i\bar{q}_{2}\cdot b}\ \hat{\delta}^{(4)}(\bq_{1} +\bq_{2} -\bar{k})\cr
		& \hspace{1cm}\big(\hbar^{2}\mathcal{A}_{5}^{\mu\nu}(p_{1}+\hbar\bar{q}_{1},p_{2}+\hbar\bar{q}_{2}\rightarrow p_{1},p_{2},\hbar\bar{k})\big)\Big \rrangle \,.
	\end{align}
Here $\Big \llangle f(p_{1},p_{2},q\ldots)\Big \rrangle$ denotes the integration over the minimum uncertainty wave packets which localizes the momenta onto their classical values. Note that in the above expression, we have stripped the $\hbar-$ scaling of the coupling constant.
The classical limit of the radiation kernel is the radiative gravitational field. To obtain the soft radiative field, we then carry out a soft expansion of the radiation kernel in the frequency of the emitted graviton. This will give us the classical radiation to (sub)$^n$-leading order in frequency. 
	The KMOC formalism has been generalized to describe many types of scattering. For example, in \cite{Cristofoli:2021vyo} the formalism has been extended to include incoming waves in the initial state. It has also been extended to include additional internal degrees of freedom, such as color charges in \cite{delaCruz:2020bbn}. It has also been generalized to describe scattering in curved backgrounds \cite{Adamo:2021rfq, Adamo:2022rmp}. 

\section{A brief review of classical log soft theorems} \label{sec:classicallogs}
Classical soft theorems govern the non-analytic behavior of electromagnetic and gravitational waveforms in the low-frequency domain of the detector. The most well-known example of the classical soft graviton theorem is the gravitational memory effect \cite{Zeldovich:1974gvh,Christodoulou:1991cr,Wiseman:1991ss,Thorne:1992sdb} which predicts a permanent change in the asymptotic metric fluctuation caused by the passage of a gravitational wave. Memory effect is an observable of classical scattering and is simply the coefficient of the leading term in the soft expansion of the radiative field \cite{Weinberg:1964ew,Weinberg:1965nx,Strominger:2014pwa}. As reviewed in the introduction, in recent years, a hierarchy of universal soft theorems in four dimensions have been discovered \cite{Laddha:2018myi,Sahoo:2018lxl,Saha:2019tub,Sahoo:2020ryf,Sahoo:2021ctw,Ghosh:2021bam}.

The table in \cite{Ghosh:2021bam} summarizes the different orders of the low-frequency gravitational waveform in the $\omega \rightarrow 0$ limit and their relations to PM expansion. It also illustrates the late and early time behavior of the gravitational waveform at large retarded time $u$.
In this paper, we focus on tree-level scattering. The loop-level corrections to the leading logs will be pursued elsewhere. \\
As stated before, soft graviton theorems are exact statements describing the gravitational radiation in soft frequency expansion. We consider a $2 \rightarrow 2$ scattering
process with a large impact parameter. These processes can then be studied within perturbation
theory. If $p_a^\prime$ 
is the final momentum of a particle and the initial momentum is $p_a$, then we have
\begin{align} \label{perturb}
    p_a^{\prime \mu}  = p_a^\mu + \sum_{n=1}^\infty \kappa^{2n} \Delta p_a^{(n) \mu} \,,
\end{align}
where $\Delta p_a^{(1) \mu}$ is the LO linear impulse and $\kappa^{2n}$-th term is the N$^{n-1}$LO impulse. Consistency requires that the radiation at any perturbative order matches the classical soft factor. Plugging the equation \eqref{perturb} in the radiative field \eqref{softexp}, it is evident that both $\log{\omega}$ and $\omega\log{\omega}$ survive even at leading order in the coupling. This then connects the appearance of $\omega^{n}\log{\omega}$ from tree-level amplitudes which we will compute in the subsequent sections. 

For example, in 2-2 scattering the classical log soft graviton factor is given by
\begin{align}
    h^{\mu\nu} (\omega,\hat{n})|_{\log{\omega}} &= \frac{\kappa^3}{16 \pi} \log{ (\omega )} \left(\sum_{a,b=1}^2 S^{(1),\mu\nu}(\{p_a\},k) +  \sum_{a,b=1}^2 S^{(1),\mu\nu}(\{p_a^\prime \},k) \right) \,,
\end{align}
where
\begin{align} 
     S^{(1),\mu\nu}(\{p_a\},k) &= (p_1 \cdot p_2)  \frac{(2(p_1 \cdot p_2)^2 - 3m_1^2 m_2^2)}{[(p_1 \cdot p_2)^2 -m_1^2 m_2^2]^{3/2}} \frac{k_\rho }{p_a \cdot k} p_a^{(\mu}\left( p_a^{\nu)} p_b^\rho - p_b^{\nu)} p_a^\rho \right) \,.
\end{align}
Using equation \eqref{perturb}, the classical log soft graviton factor at leading order in the coupling takes the following form
\begin{align}
    h^{\mu\nu} (\omega,\hat{n})|_{\log{\omega}} &= -\frac{i\kappa}{4} \log{ (\omega )} \sum_i \frac{1}{p_i \cdot k} p_i^{(\mu} \hat{J}_i^{\nu)\rho} k_\rho \left[\frac{\kappa^2}{2\pi} \frac{(p_1 \cdot p_2)^2 - \frac{1}{2}m_1^2 m_2^2}{\sqrt{(p_1 \cdot p_2)^2 -m_1^2 m_2^2}}\right] \,,
\end{align}
which arises from the action on gravitational tree-level four-point amplitude. Here $\hat{J}_i^{\mu\nu} = i(p_i \wedge \frac{\partial}{\partial p_i})^{\mu\nu}$. Thus one can use the soft graviton theorems for tree-level amplitudes in computing the radiation kernel, which generates the $\omega^n \log{\omega}$ terms.

 \section{A brief review of soft graviton theorems for tree-level amplitude} \label{sec:treesoft}
 In this section, we review the infinite hierarchy of soft graviton theorems for tree-level amplitudes. We consider massive scalars coupled to gravity and analyze five-point scattering amplitude in which two scalar particles with momenta $\tilde{p}_{1}, \tilde{p}_{2}$ scatter into $p_{1}, p_{2}$ and a graviton. Our kinematics satisfy the following on-shell conditions
 \begin{align}
 p_{i}^{2} = \tilde{p}_{i}^{2} = m_{i}^{2} \,.
 \end{align}
 For a generic theory of gravity with arbitrary matter coupling,  the (sub)$^{n} \vert n \geq\ 1$ soft graviton theorem can be written as follows \cite{Hamada:2018vrw,Li:2018gnc}.
 \begin{align}
\lim_{\omega\, \rightarrow\, 0} \partial_{\omega}^n \omega{\cal A}_{5}(\tilde{p}_{1},\tilde{p}_{2} \rightarrow\, p_{1},p_{2},k)\, =\, \hat{S}^{n} {\cal A}_{4} + {\cal B}_{n}(p_{1}, p_{2}, p_{1}^{\prime}, p_{2}^{\prime}, \hat{k})\, {\cal A}_{4} + {\cal R}_{n}(p_{1}, p_{2}, p_{1}^{\prime}, p_{2}^{\prime}, \hat{k}) 
 \end{align}
where the (sub)$^{n}$-leading soft factor is defined as
 \begin{equation}
\hat{S}^{n} = \begin{cases} 
i \frac{\kappa}{2}\sum_{i=1,2} \epsilon_{\mu\nu} 
\left[\frac{1}{p_i \cdot k} p_i^{(\mu} \hat{J}_i^{\nu)\rho}k_\rho - \frac{1}{\tilde{p}_i \cdot k} \tilde{p}_i^{(\mu} \hat{\tilde{J}}_i^{\nu)\rho}k_\rho \right] \,, ~~~~~ \text{if} ~~\, n =1 \\
\frac{\kappa}{2}\sum_{i=1,2} \epsilon_{\mu\nu} \left[\frac{\hat{J}_i^{\mu\rho}k_\rho \hat{J}_i^{\nu\sigma} k_\sigma}{p_i \cdot k} \left(k \cdot \frac{\partial}{\partial p_i} \right)^{n-2} + \frac{\hat{\tilde{J}}_i^{\mu\rho}k_\rho \hat{\tilde{J}}_i^{\nu\sigma} k_\sigma}{\tilde{p}_i \cdot k}\left(k \cdot \frac{\partial}{\partial \tilde{p}_i} \right)^{n-2} \right]  ~~~~~ \text{if} ~~\, n \geq 2 \,.
\end{cases}
 \end{equation}
 ${\cal B}_{n}$ is the non-universal part of the factorization formula which depends on the irrelevant three-point couplings in the theory \cite{Elvang:2016qvq,Laddha:2017ygw}. 
\begin{align} \label{antisym}
    {\cal R}_{n} = \epsilon_{\mu\nu}\hat{k}_{\alpha_1} \hat{k}_{\alpha_2} \cdots \hat{k}_{\alpha_{n-1}} A^{\mu\nu \alpha_1 \alpha_2 \cdots \alpha_{n-1}} 
\end{align}
is the so called Remainder term which vanishes for $n < 3$ and spoils factorisation for $n \geq 3$. $A$ is antisymmetric in any two
indices among $\mu$ and $\alpha$’s. As was shown in \cite{Campiglia:2018dyi}, however, the Remainder term always satisfies the following constraint simply due to the tensor structure of the amplitude.
\begin{align} \label{constrainteq}
D_{z}^{n+1} (1 + |z|^2)^{-1} {\cal R}_{n} = 0 \,,
\end{align} 
where $\hat{k} = (1,\hat{n}(z,\bar{z}))$ and $z,\bar{z}$ are the stereographic coordinates.
  In the present paper, we consider the gravitational scattering of two minimally coupled scalars $\phi_{1}, \phi_{2}$ with masses $m_{1}, m_{2}$ respectively, and as a result 
  \begin{align}
{\cal B}_{n} = 0\, \forall \, n
\end{align}
  However, as discussed above, the remainder term ${\cal R}_{n}$ continue to persist $\forall\, n\, \geq\, 3$.\\

\begin{figure}[h!]
	\centering
	\begin{tikzpicture}[scale=1,photon/.style={decorate,decoration={snake, amplitude=0.5mm,segment length=1.75mm}}]
		\draw[thick] (-1,-1) -- (0,0) -- (-1,1);
		\draw[photon, thick, red] (0,0) -- (1.5,0);
		\draw[->,photon, thick] (2,0.5) -- (1.5,1);
		\draw[thick] (2.5,-1) -- (1.5,0) -- (2.5,1);
		\draw[-latex, thick] (-0.5,-0.5) -- (-0.51,-0.51);
		\draw[-latex, thick] (-0.5,0.5) -- (-0.49,0.49);
		\draw[-latex, thick] (2.2,0.7) -- (2.19,0.69);
		\draw[-latex, thick] (2,-0.5) -- (2.01,-0.51);
		\draw[-latex] (0.25,-0.2) -- (1.25,-0.2) ;
		\node[red] at (0.75,-0.35){$q_2$};
		\node[red] at (-1.2,-1.2) {$\mathbf{2}$};
		\node[red] at (-1.2,1.2) {$\mathbf{\bar{2}}$};
		\node[red] at (2.7,-1.2) {$\mathbf{1}$};
		\node[red] at (2.7,1.2) {$\mathbf{\bar{1}}$};
		\node[red] at (1.4,1.2) {$k$};
		\node[] at (1,-1.5) {I};
	\end{tikzpicture} 
	\begin{tikzpicture}[scale=1,photon/.style={decorate,decoration={snake, amplitude=0.5mm,segment length=1.75mm}}]
		\draw[thick] (-1,-1) -- (0,0) -- (-1,1);
		\draw[photon, thick, red] (0,0) -- (1.5,0);
		\draw[->,photon, thick] (2,-0.5) -- (2.5,0);
		\draw[thick] (2.5,-1) -- (1.5,0) -- (2.5,1);
		\draw[-latex, thick] (-0.5,-0.5) -- (-0.51,-0.51);
		\draw[-latex, thick] (-0.5,0.5) -- (-0.49,0.49);
		\draw[-latex, thick] (2.01,0.51) -- (2,0.5);
		\draw[-latex, thick] (2.3,-0.8) -- (2.31,-0.81);
		\draw[-latex] (0.25,-0.2) -- (1.25,-0.2) ;
		\node[red] at (0.75,-0.35){$q_2$};
		\node[red] at (-1.2,-1.2) {$\mathbf{2}$};
		\node[red] at (-1.2,1.2) {$\mathbf{\bar{2}}$};
		\node[red] at (2.7,-1.2) {$\mathbf{1}$};
		\node[red] at (2.7,1.2) {$\mathbf{\bar{1}}$};
		\node[red] at (2.7,0.1) {$k$};
		\node[] at (1,-1.5) {II};
	\end{tikzpicture}
	\begin{tikzpicture}[scale=1,photon/.style={decorate,decoration={snake, amplitude=0.5mm,segment length=1.75mm}}]
		\draw[thick] (-1,-1) -- (0,0) -- (-1,1);
		\draw[photon, thick, red] (0,0) -- (1.5,0);
		\draw[->,photon, thick] (1.5,0) -- (1,1);
		\draw[thick] (2.5,-1) -- (1.5,0) -- (2.5,1);
		\draw[-latex, thick] (-0.5,-0.5) -- (-0.51,-0.51);
		\draw[-latex, thick] (-0.5,0.5) -- (-0.49,0.49);
		\draw[-latex, thick] (2.01,0.51) -- (2,0.5);
		\draw[-latex, thick] (2,-0.5) -- (2.01,-0.51);
		\draw[-latex] (0.25,-0.2) -- (1.25,-0.2) ;
		\node[red] at (0.75,-0.35){$q_2$};
		\node[red] at (-1.2,-1.2) {$\mathbf{2}$};
		\node[red] at (-1.2,1.2) {$\mathbf{\bar{2}}$};
		\node[red] at (2.7,-1.2) {$\mathbf{1}$};
		\node[red] at (2.7,1.2) {$\mathbf{\bar{1}}$};
		\node[red] at (0.9,1.2) {$k$};
		\node[] at (1,-1.5) {III};
	\end{tikzpicture}
 \begin{tikzpicture}[scale=1,photon/.style={decorate,decoration={snake, amplitude=0.5mm,segment length=1.75mm}}]
		\draw[thick] (-1,-1) -- (0,0) -- (-1,1);
		\draw[photon, thick, red] (0,0) -- (1.5,0);
		\draw[->,photon, thick] (-0.5,0.5) -- (0,1);
		\draw[thick] (2.5,-1) -- (1.5,0) -- (2.5,1);
		\draw[-latex, thick] (-0.5,-0.5) -- (-0.51,-0.51);
		\draw[-latex, thick] (-0.6,0.6) -- (-0.59,0.59);
		\draw[-latex, thick] (2.01,0.51) -- (2,0.5);
		\draw[-latex, thick] (2,-0.5) -- (2.01,-0.51);
		\draw[-latex] (1.25,-0.2) -- (0.25,-0.2) ;
		\node[red] at (0.75,-0.35){$q_1$};
		\node[red] at (-1.2,-1.2) {$\mathbf{2}$};
		\node[red] at (-1.2,1.2) {$\mathbf{\bar{2}}$};
		\node[red] at (2.7,-1.2) {$\mathbf{1}$};
		\node[red] at (2.7,1.2) {$\mathbf{\bar{1}}$};
		\node[red] at (0.2,1.2) {$k$};
		\node[] at (1,-1.5) {IV};
	\end{tikzpicture}
	\begin{tikzpicture}[scale=1,photon/.style={decorate,decoration={snake, amplitude=0.5mm,segment length=1.75mm}}]
		\draw[thick] (-1,-1) -- (0,0) -- (-1,1);
		\draw[photon, thick, red] (0,0) -- (1.5,0);
		\draw[->,photon, thick] (-0.5,-0.5) -- (-1,0);
		\draw[thick] (2.5,-1) -- (1.5,0) -- (2.5,1);
		\draw[-latex, thick] (-0.8,-0.8) -- (-0.81,-0.81);
		\draw[-latex, thick] (-0.5,0.5) -- (-0.49,0.49);
		\draw[-latex, thick] (2.01,0.51) -- (2,0.5);
		\draw[-latex, thick] (2,-0.5) -- (2.01,-0.51);
		\draw[-latex] (1.25,-0.2) --(0.25,-0.2)  ;
		\node[red] at (0.75,-0.35){$q_1$};
		\node[red] at (-1.2,-1.2) {$\mathbf{2}$};
		\node[red] at (-1.2,1.2) {$\mathbf{\bar{2}}$};
		\node[red] at (2.7,-1.2) {$\mathbf{1}$};
		\node[red] at (2.7,1.2) {$\mathbf{\bar{1}}$};
		\node[red] at (-1.1,0.2) {$k$};
		\node[] at (1,-1.5) {V};
	\end{tikzpicture}
	\begin{tikzpicture}[scale=1,photon/.style={decorate,decoration={snake, amplitude=0.5mm,segment length=1.75mm}}]
		\draw[thick] (-1,-1) -- (0,0) -- (-1,1);
		\draw[photon, thick, red] (0,0) -- (1.5,0);
		\draw[->,photon, thick] (0,0) -- (1,1);
		\draw[thick] (2.5,-1) -- (1.5,0) -- (2.5,1);
		\draw[-latex, thick] (-0.5,-0.5) -- (-0.51,-0.51);
		\draw[-latex, thick] (-0.5,0.5) -- (-0.49,0.49);
		\draw[-latex, thick] (2.01,0.51) -- (2,0.5);
		\draw[-latex, thick] (2,-0.5) -- (2.01,-0.51);
		\draw[-latex] (1.25,-0.2) -- (0.25,-0.2) ;
		\node[red] at (0.75,-0.35){$q_1$};
		\node[red] at (-1.2,-1.2) {$\mathbf{2}$};
		\node[red] at (-1.2,1.2) {$\mathbf{\bar{2}}$};
		\node[red] at (2.7,-1.2) {$\mathbf{1}$};
		\node[red] at (2.7,1.2) {$\mathbf{\bar{1}}$};
		\node[red] at (1.1,1.2) {$k$};
		\node[] at (1,-1.5) {VI};
	\end{tikzpicture}
 \begin{tikzpicture}[scale=1,photon/.style={decorate,decoration={snake, amplitude=0.5mm,segment length=1.75mm}}]
		\draw[thick] (-1,-1) -- (0,0) -- (-1,1);
		\draw[photon, thick, red] (0,0) -- (1.5,0);
		\draw[->,photon, thick] (0.75,0) -- (0.75,1);
		\draw[thick] (2.5,-1) -- (1.5,0) -- (2.5,1);
		\draw[-latex, thick] (-0.5,-0.5) -- (-0.51,-0.51);
		\draw[-latex, thick] (-0.5,0.5) -- (-0.49,0.49);
		\draw[-latex, thick] (2.01,0.51) -- (2,0.5);
		\draw[-latex, thick] (2,-0.5) -- (2.01,-0.51);
		\draw[-latex] (1.4,-0.2) -- (0.9,-0.2) ;
            \draw[-latex] (0.1,-0.2) -- (0.6,-0.2) ;
		\node[red] at (1.2,-0.4){$q_1$};
            \node[red] at (0.3,-0.4){$q_2$};
		\node[red] at (-1.2,-1.2) {$\mathbf{2}$};
		\node[red] at (-1.2,1.2) {$\mathbf{\bar{2}}$};
		\node[red] at (2.7,-1.2) {$\mathbf{1}$};
		\node[red] at (2.7,1.2) {$\mathbf{\bar{1}}$};
		\node[red] at (1,1.1) {$k$};
		\node[] at (1,-1.5) {VII};
	\end{tikzpicture}
	\caption{Tree-level five-point amplitudes for gravitational scattering of massive particles}
\label{7b}
\end{figure}
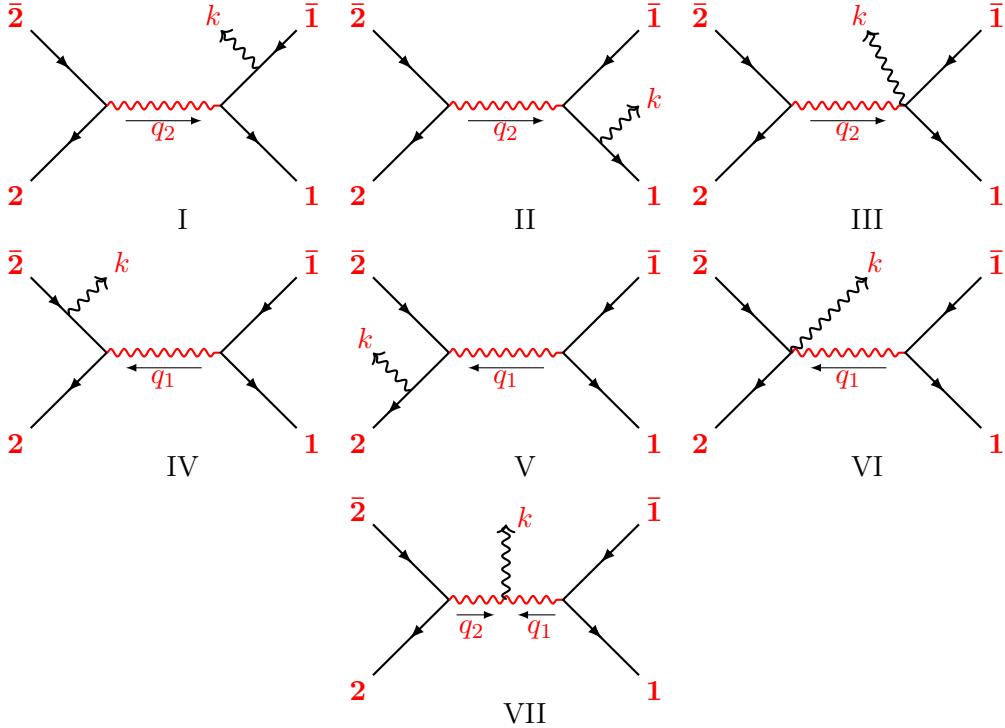

The leading order contribution to the gravitational radiation arises from the tree-level five-point amplitude which in this case is a sum over seven Feynman diagrams shown in Figure \ref{7b}.

The unstripped five-point amplitude is given by
\begin{align}
    \bar{\mathcal{M}}_{5}^{\mu\nu}[p_{1}+\hbar\bar{q}_{1},p_{2}+\hbar\bar{q}_{2}\rightarrow p_{1},p_{2,},\hbar\bar{k}] = \hat{\delta}^{(4)}(\bq_{1} +\bq_{2} -\bar{k}) \bar{\mathcal{A}}_{5}^{\mu\nu}[p_{1}+\hbar\bar{q}_{1},p_{2}+\hbar\bar{q}_{2}\rightarrow p_{1},p_{2},\hbar\bar{k}] \,.
\end{align}
Here the stripped amplitude $\bar{\mathcal{A}}_{5}^{\mu\nu}$ is given by \cite{Luna:2017dtq,Bohnenblust:2023qmy}:
\begin{align} \label{5pointstrip}
\bar{\mathcal{A}}_{5}^{\mu\nu}[p_{1}+\hbar\bar{q}_{1},p_{2}+\hbar\bar{q}_{2}\rightarrow p_{1},p_{2,},\hbar\bar{k}] = -\frac{\kappa^3 m_1^2 m_2^2}{\hbar^2 }\Big[ \frac{4  P^\mu P^\nu}{\bq_1^2 \bq_2^2} &+ \frac{2\gamma}{\bq_1^2 \bq_2^2}  \Big(Q^\mu P^\nu + Q^\nu P^\mu  \Big) \cr
 &+ \Big(\gamma^2 - \frac{1}{2} \Big)\Big(\frac{Q^\mu Q^\nu}{\bq_1^2 \bq_2^2} - \frac{P^\mu P^\nu}{\omega_1^2 \omega_2^2} \Big) \Big] \,, \cr
\end{align}
where $\kappa = \sqrt{32\pi G}$, $\gamma= u_1 \cdot u_2$ and
\begin{align}
    P^\mu &= -\omega_1 u_2^\mu + \omega_2 u_1^\mu \cr
    Q^\mu &= (\bq_1 - \bq_2)^\mu + \frac{\bq_1^2}{\omega_1}u_1^\mu - \frac{\bq_2^2}{\omega_2}u_2^\mu \,, \ \omega_1 = - \bar{k} \cdot u_1 , \ \omega_2 = - \bar{k} \cdot u_2 \,.
\end{align}
As $k_\mu \bar{\mathcal{A}}_{5}^{\mu\nu} = k_\nu \bar{\mathcal{A}}_{5}^{\mu\nu} = 0$, the amplitude is gauge-invariant. The (sub)$^{n}$-leading soft graviton theorem for the tree-level amplitude can be restated as follows
\begin{align} \label{factoramp}
    &\hat{\delta}^{(4)}(q_{1} + q_{2} -k) \mathcal{A}_{5}^{\mu\nu}[p_{1}+q_{1},p_{2}+q_{2}\rightarrow p_{1},p_{2},k] \cr
    &= \sum_{r=0}^{n} \frac{(-1)^{n-r}}{(n-r)!} S^{(r),\mu\nu} (k \cdot \partial)^{n-r} (\hat{\delta}^{(4)}(q_{1} +q_{2}))\mathcal{A}_{4}[p_{1}+q_{1},p_{2}+q_{2}\rightarrow p_{1},p_{2}]  + \mathcal{X}^{\mu\nu} \,,
\end{align}
where the soft graviton factors are given by
\begin{align} \label{gravitonsoftfactor}
     & S^{(0),\mu\nu} = \kappa \sum_{i=1,2} \left[\frac{1}{p_i \cdot k} p_i^{(\mu} p_i^{\nu)} - \frac{1}{\tilde{p}_i \cdot k} \tilde{p}_i^{(\mu} \tilde{p}_i^{\nu)} \right] \cr
     & S^{(1),\mu\nu} = i \frac{\kappa}{2}\sum_{i=1,2} \left[\frac{1}{p_i \cdot k} p_i^{(\mu} \hat{J}_i^{\nu)\rho}k_\rho - \frac{1}{\tilde{p}_i \cdot k} \tilde{p}_i^{(\mu} \hat{\tilde{J}}_i^{\nu)\rho}k_\rho \right] \cr
      & S^{(n),\mu\nu} = \frac{\kappa}{2}\sum_{i=1,2} \left[\frac{\hat{J}_i^{\mu\rho}k_\rho \hat{J}_i^{\nu\sigma} k_\sigma}{p_i \cdot k} \left(k \cdot \frac{\partial}{\partial p_i} \right)^{n-2} + \frac{\hat{\tilde{J}}_i^{\mu\rho}k_\rho \hat{\tilde{J}}_i^{\nu\sigma} k_\sigma}{\tilde{p}_i \cdot k}\left(k \cdot \frac{\partial}{\partial \tilde{p}_i} \right)^{n-2} \right]  \  \,, n \geq 2  \,. \cr
\end{align}
$\mathcal{R}_n = \epsilon_{\mu\nu}\mathcal{X}^{\mu\nu}$ is the remainder term that spoils the factorization beyond the sub-subleading order in the soft expansion. For the amplitude \eqref{5pointstrip}, it is given by the following expression
\begin{align} \label{remainderamp}
    \mathcal{X}^{\mu\nu} = \frac{\kappa^3 m_1^2 m_2^2}{4} \sum_{r=3}^{n} \frac{(-1)^{n-r}}{(n-r)!} (k \cdot \partial)^{n-r} (\hat{\delta}^{(4)}(q_{1} +q_{2}))  \Lambda_{r-1}^{\mu\nu} + (1 \leftrightarrow 2)\,,
\end{align}
where the polynomial $ \Lambda_{n}^{\mu\nu}$ is defined as
\begin{align} \label{Lpolynomial}
    \Lambda_{n \geq 2}^{\mu\nu} &=  H_2^{\mu\nu} \frac{2^{n-2}(\bq \cdot \bar{k})^{n-2}}{(\bq^2)^{n-2}} \,.
\end{align}
Here
\begin{align} 
    H_2^{\mu\nu} = -\frac{4}{(\bq^2)^2} \Big( \omega_2^2 u_1^\mu u_1^\nu - \frac{\omega_1 \omega_2}{2}(u_2^\mu u_1^\nu + u_2^\nu u_1^\mu)\Big) \,.
\end{align}
As can be explicitly checked ${\cal R}_{n}$ satisfies the constraint equation \eqref{constrainteq} and the resulting $A$ is antisymmetric in the indices as stated in equation \eqref{antisym}. 
The factorized terms that were obtained via the soft graviton factors on the four-point amplitude in equation \eqref{factoramp} can be 
written as follows
\begin{align} \label{factorized}
    \frac{\kappa^3 m_1^2 m_2^2}{4} \sum_{r=0}^{n} \frac{(-1)^{n-r}}{(n-r)!} (k \cdot \partial)^{n-r} (\hat{\delta}^{(4)}(q_{1} +q_{2}))  K_{r-1}^{\mu\nu} + (1 \leftrightarrow 2)\,,
\end{align}
where the polynomials $K_{n}^{\mu\nu}$ are defined as 
\begin{align} \label{Kpolynomials}
    K_{-1}^{\mu\nu} &= -\frac{1}{\bq^2}\Big(\gamma^2 - \frac{1}{2} \Big)\Big\{- \frac{1}{\omega_1} \bq^{(\mu} u_1^{\nu)}  - \frac{1}{\omega_1^2}(\bar{k} \cdot \bq) u_1^\mu u_1^\nu  \Big\} \cr
    K_{0}^{\mu\nu} &= -\frac{2\gamma}{ \bq^2} \Big(-  u_1^{(\mu} u_2^{\nu)} + \frac{2\omega_2}{\omega_1} u_1^\mu u_1^\nu  \Big) + H_0^{\mu\nu} \cr
    K_{+1}^{\mu\nu} &= H_1^{\mu\nu} + H_0^{\mu\nu}  \frac{2(\bq \cdot \bar{k})}{\bq^2} \cr
    K_{n \geq 2}^{\mu\nu} &= H_1^{\mu\nu}  \frac{2^{n-1}(\bq \cdot \bar{k})^{n-1}}{(\bq^2)^{n-1}} + H_0^{\mu\nu} \frac{2^{n}(\bq \cdot \bar{k})^{n}}{(\bq^2)^{n}} \,,
\end{align}
where
\begin{align}
    H_1^{\mu\nu} = \frac{4\gamma}{\bq^2} \frac{\omega_2}{\bq^2} \bq^{(\mu} u_1^{\nu)} ~~~~~\text{and} \, \ H_0^{\mu\nu} = -\frac{2}{\bq^2}\Big(\gamma^2 - \frac{1}{2} \Big)\frac{\bq^\mu \bq^\nu}{\bq^2} 
\end{align}

 \section{Gravitational scattering of massive spinless particles} \label{sgravitysec}
We consider gravitational scattering of two scalars $\phi_{1}, \phi_{2}$ with masses $m_{1}, m_{2}$ respectively. KMOC formalism provides us with a formula to compute the gravitational radiative field. In a scattering process where two scalar particles with initial momenta $p_{1}, p_{2}$ gravitationally scatter at impact parameter distance $b$, the radiative field is given by the following equation 
\begin{align}
h_{\mu\nu}(\omega, \hat{n})\, &=\, \lim_{\hbar\, \rightarrow\, 0}\, \hbar^{3/2}\, \int \prod_{i=1}^{2} \hat{d}^{4}\bar{q}_{i}\  \hat{\delta}(2p_{i}\cdot \bar{q}_{i}) \ e^{i\bar{q}_{1}\cdot b}\ \hat{\delta}^{(4)}(\bq_{1} +\bq_{2} -\bar{k})\cr
		&\hspace{3cm}\big(\mathcal{A}_{5}^{\mu\nu}(p_{1}+\hbar\bar{q}_{1},p_{2}+\hbar\bar{q}_{2}\rightarrow p_{1},p_{2},\hbar\bar{k}\big) + \cdots + \mathcal{O}(\kappa^{5})
\end{align}
where $\mathcal{A}_{5}^{\mu\nu}$ is the five point amplitude. The final momenta are integrated over and this integration is written in terms of the momentum exchange $q_{i} = \tilde{p}_{i} - p_{i}$. The leading order contribution to the gravitational radiation arises from the tree-level amplitude \eqref{5pointstrip} and it is given by
\begin{align} \label{exactkernel}
\mathcal{R}^{\mu\nu}(\omega,\hat{n}) &=  -\frac{\kappa^3 m_1 m_2}{4}  \int \hat{d}^4 \bar{q} \hat{\delta}(u_1 \cdot \bar{k} - u_1 \cdot \bar{q}) \hat{\delta}(u_2 \cdot \bar{q}) e^{i b\cdot (\bar{k} - \bar{q})}  \cr 
&\qquad \times \Big[ \frac{4}{(\bar{k} - \bq)^2 \bq^2} \Big( \omega_2^2 u_1^\mu u_1^\nu - \frac{\omega_1 \omega_2}{2}(u_2^\mu u_1^\nu + u_2^\nu u_1^\mu)\Big) \cr
    &\qquad + \frac{2\gamma}{ \bq^2} \Big(-2\frac{\omega_2}{(\bar{k} - \bq)^2} \bq^{(\mu}u_1^{\nu)} -  u_1^{(\mu} u_2^{\nu)} + \frac{2\omega_2}{\omega_1} u_1^\mu u_1^\nu  \Big) \cr
    &\qquad + \frac{2}{\bq^2}\Big(\gamma^2 - \frac{1}{2} \Big)\Big\{\frac{1}{(\bar{k} - \bq)^2} \bq^\mu \bq^\nu - \frac{1}{\omega_1} \bq^{(\mu} u_1^{\nu)}  - \frac{1}{\omega_1^2}(\bar{k} \cdot \bq) u_1^\mu u_1^\nu  \Big\} \Big] \cr
& -\frac{\kappa^3 m_1 m_2}{4} \int \hat{d}^4 \bar{q} \hat{\delta}(u_1 \cdot \bar{q}) \hat{\delta}(u_2 \cdot \bar{k}- u_2 \cdot \bar{q}) e^{i b\cdot \bar{q}} \cr 
&\qquad \times \Big[ \frac{4}{(\bar{k} - \bq)^2 \bq^2} \Big( \omega_1^2 u_2^\mu u_2^\nu - \frac{\omega_1 \omega_2}{2}(u_2^\mu u_1^\nu + u_2^\nu u_1^\mu)\Big) \cr
    &\qquad + \frac{2\gamma}{ \bq^2} \Big(-2\frac{\omega_1}{(\bar{k} - \bq)^2} \bq^{(\mu}u_2^{\nu)} -  u_2^{(\mu} u_1^{\nu)} + \frac{2\omega_1}{\omega_2} u_2^\mu u_2^\nu  \Big) \cr
    &\qquad + \frac{2}{\bq^2}\Big(\gamma^2 - \frac{1}{2} \Big)\Big\{\frac{1}{(\bar{k} - \bq)^2} \bq^\mu \bq^\nu - \frac{1}{\omega_2} \bq^{(\mu} u_2^{\nu)}  - \frac{1}{\omega_2^2}(\bar{k} \cdot \bq) u_2^\mu u_2^\nu  \Big\} \Big]  \,. 
\end{align}
\\
Soft expansion of the radiation kernel is obtained via the following two expansions:
\begin{align} \label{deltaexp}
    \hat{\delta}(u_1 \cdot \bar{k} - u_1 \cdot \bar{q}) &= \hat{\delta}(u_1 \cdot \bar{q}) - (u_1 \cdot \bar{k})\hat{\delta}^\prime(u_1 \cdot \bar{q}) + \frac{(u_1 \cdot \bar{k})^2}{2!} \hat{\delta}^{\prime\prime}(u_1 \cdot \bar{q}) \cr
    &\hspace{4cm}+ \cdots + (-1)^n \frac{(u_1 \cdot \bar{k})^n}{n!} \hat{\delta}^{(n)}(u_1 \cdot \bar{q})
\end{align}
and
\begin{align} \label{taylorexp}
    \frac{1}{(\bar{k}- \bq)^2} &=  \frac{1}{\bq^2 - 2(\bq \cdot \bar{k})} = \frac{1}{\bq^2} \Big(1 +  \frac{2(\bq \cdot \bar{k})}{\bq^2} +  \frac{4(\bq \cdot \bar{k})^2}{(\bq^2)^2} + \frac{8(\bq \cdot \bar{k})^3}{(\bq^2)^3}  + \cdots  \Big)
\end{align}
Note that the range of integration over exchange momentum is $\omega < |q| < b^{-1}$, where $\omega = \vert k\vert$ and hence the integral of the terms from the expansion in equation \eqref{taylorexp} is IR finite. The integration region is naturally restricted to $|q| \geq \omega$. One way to understand this is to notice that the soft expansion of the unstripped amplitude is obtained by taylor expansion of the delta function shown above which implicitly assumes that $\omega \ll |q|$. The upper limit of the integration region in the soft limit is automatically imposed by the phase term in the integrand as the scattering length is given by $1/|q| \sim b$.
The leading order soft radiation can be written as follows
\begin{align}
    \mathcal{R}^{(0),\mu\nu} &= \frac{\kappa^3 m_1 m_2}{4}  \int \hat{d}^4 \bar{q} \Big\{e^{-i b\cdot \bar{q}} \hat{\delta}(u_1 \cdot \bar{q}) \hat{\delta}(u_2 \cdot \bar{q}) \Big( K_{-1}^{\mu\nu} \Big) + e^{i b\cdot \bar{q}} \Big( 1 \leftrightarrow 2 \Big) \Big\} \cr
    &= \frac{\kappa^3 m_1 m_2}{4}  \Big(\gamma^2 - \frac{1}{2} \Big) \int \frac{\hat{d}^4 \bar{q}}{\bq^2} e^{-i b\cdot \bar{q}} \hat{\delta}(u_1 \cdot \bar{q}) \hat{\delta}(u_2 \cdot \bar{q})\Big( \frac{1}{\omega_1} \bq^{(\mu} u_1^{\nu)}   + \frac{(\bar{k} \cdot \bq)}{\omega_1^2} u_1^\mu u_1^\nu  \Big)  + (1 \leftrightarrow 2) \cr
    &= -\frac{\kappa^3 m_1 m_2}{8\pi \gamma\beta b^2} \Big(\gamma^2 - \frac{1}{2} \Big) \Big( \frac{1}{\omega_1} b^{(\mu} u_1^{\nu)}   + \frac{(\bar{k} \cdot b)}{\omega_1^2} u_1^\mu u_1^\nu  \Big)  +  (1 \leftrightarrow 2)\,.
\end{align}
Note that in the $\vert b\vert \rightarrow\, \infty$ limit, this term doesn't contribute to the kernel. This is expected as the particles are so far apart in this limit that they experience no deflection, and hence the memory effect vanishes! However, we show below that, unlike the memory effect, the leading log soft factor survives in this vanishing deflection limit.

Using the expansion of the delta functions and the five-point amplitude, the sub-leading order soft radiation can be expressed as follows
\begin{align}
    \mathcal{R}^{(1),\mu\nu} &= \frac{\kappa^3 m_1 m_2}{4}  \int \hat{d}^4 \bar{q} \Big[\Big\{e^{-i b\cdot \bar{q}} \hat{\delta}(u_1 \cdot \bar{q}) \hat{\delta}(u_2 \cdot \bar{q}) \Big( K_{0}^{\mu\nu} \Big) + e^{i b\cdot \bar{q}} \Big( 1 \leftrightarrow 2 \Big) \Big\}\cr
&\hspace{3cm} - \Big\{e^{-i b\cdot \bar{q}} \hat{\delta}^\prime(u_1 \cdot \bar{q}) \hat{\delta}(u_2 \cdot \bar{q}) (u_1 \cdot k)\Big( K_{-1}^{\mu\nu}\Big) + e^{i b\cdot \bar{q}} \Big( 1 \leftrightarrow 2 \Big) \Big\} \cr
&\hspace{3cm}+ e^{-i b\cdot \bar{q}} \hat{\delta}(u_1 \cdot \bar{q}) \hat{\delta}(u_2 \cdot \bar{q}) (i b \cdot \bar{k}) K_{-1}^{\mu\nu}  \Big] \,.
\end{align}
Here, the superscript in $ \mathcal{R}^{(1),\mu\nu}$ denotes the truncation of the radiation kernel to $\mathcal{O}(\omega^0)$.

We can re-write $\mathcal{R}^{(1),\mu\nu}$ as
\begin{align}
\mathcal{R}^{(1),\mu\nu} = \frac{\kappa^3 m_1 m_2}{4} \left( I_{1}^{\mu\nu} + I_{2}^{\mu\nu} + I_{3}^{\mu\nu} \right)
\end{align}
All three terms are evaluated in Appendix \ref{allorderintegrals} and the final result is summarised below.  
\begin{align}
    I_1^{\mu\nu} &= \int \hat{d}^4 \bar{q} e^{-i b\cdot \bar{q}} \hat{\delta} (u_1 \cdot \bar{q}) \hat{\delta}(u_2 \cdot \bar{q}) \Big[\frac{2\gamma}{ \bq^2} \Big(-  u_1^{(\mu} u_2^{\nu)} + \frac{2\omega_2}{\omega_1} u_1^\mu u_1^\nu  \Big) +\frac{2}{\bq^2}\Big(\gamma^2 - \frac{1}{2} \Big)\frac{\bq^\mu \bq^\nu}{\bq^2} \Big]\cr
    &= \frac{\gamma}{\pi \gamma \beta} \log{(\omega b)}  \Big(u_1^{(\mu} u_2^{\nu)} - \frac{2(u_2 \cdot \bar{k})}{(u_1 \cdot \bar{k})} u_1^\mu u_1^\nu  \Big) + O(\omega^{0}) 
\end{align}
\begin{align}
    I_2^{\mu\nu} &= -\int \frac{\hat{d}^4 \bar{q}}{\bq^2} e^{-i b\cdot \bar{q}} \hat{\delta}^\prime(u_1 \cdot \bar{q}) \hat{\delta}(u_2 \cdot \bar{q}) \Big(\gamma^2 - \frac{1}{2} \Big)(u_1 \cdot \bar{k})\Big( \frac{1}{\omega_1} \bq^{(\mu} u_1^{\nu)}   + \frac{(\bar{k} \cdot \bq)}{\omega_1^2} u_1^\mu u_1^\nu  \Big) \cr
    & = \frac{1}{2\pi \gamma^3 \beta^3}\Big(\gamma^2 - \frac{1}{2} \Big) \log (\omega b) \Big\{-(\gamma u_2 -u_1)^{(\mu} u_1^{\nu)} + \big(\gamma \frac{(u_2 \cdot \bar{k})}{u_1 \cdot \bar{k}} - 1\big) u_1^\mu u_1^\nu \Big\}\,,
\end{align}
and finally, 
\begin{align}
    I_3^{\mu\nu} &= \int \frac{\hat{d}^4 \bar{q}}{\bq^2} e^{-i b\cdot \bar{q}} \hat{\delta}(u_1 \cdot \bar{q}) \hat{\delta}(u_2 \cdot \bar{q}) (i b \cdot \bar{k}) \Big(\gamma^2 - \frac{1}{2} \Big) \Big( \frac{1}{\omega_1} \bq^{(\mu} u_1^{\nu)}   + \frac{(\bar{k} \cdot \bq)}{\omega_1^2} u_1^\mu u_1^\nu  \Big) \cr
    & = -\frac{(b \cdot k)}{2\pi \gamma\beta b^2} \Big(\gamma^2 - \frac{1}{2} \Big) \Big( \frac{1}{\omega_1} b^{(\mu} u_1^{\nu)}   + \frac{(\bar{k} \cdot b)}{\omega_1^2} u_1^\mu u_1^\nu  \Big)  \,,
\end{align}
We note that $I_3^{\mu\nu}$ has no logarithmic terms unlike $I_1^{\mu\nu}$ and $I_2^{\mu\nu}$.
Therefore the logarithmic contribution of the first particle with initial momentum $p_{1}$ to the gravitational radiation is given by 
\begin{align} \label{logsoftsub}
    \mathcal{R}_1^{\log{\omega},\mu\nu} &= \frac{\kappa^3 m_1 m_2}{4\pi \gamma^3 \beta^3} \log{ (\omega b)}  \gamma (2\gamma^2 - 3) \Big( u_1^{(\mu} u_2^{\nu)} - \frac{(u_2 \cdot \bar{k})}{(u_1 \cdot \bar{k})} u_1^{(\mu} u_1^{\nu)}  \Big) \,.
\end{align}
This expression matches with the classical result for sub-leading log soft graviton factor \cite{Laddha:2018myi,Sahoo:2018lxl} and with the result obtained using quantum soft theorems \cite{Manu:2020zxl}. As stated before, in the deflection less limit ($\vert b\vert \rightarrow\, \infty$) such that $\omega b $ is fixed, the leading logarithmic contribution survives! In fact, all the log terms of the form $\omega^{m}\log{\omega} \vert m \geq\, 1$ survive and can be completely determined by the (sub)$^{n}$-leading soft graviton theorems for tree-level gravitational amplitudes.

We can similarly compute the radiation kernel at (sub)$^{2}$-leading order in the soft expansion at leading order in the coupling given in Appendix \ref{sec:subsubsoft} and the result matches with existing results in the literature. 

Similarly, using the expansion of the delta functions and the five-point amplitude given in equations \eqref{deltaexp} and \eqref{taylorexp} respectively, the (sub)$^n$-leading order soft radiation is given by
\begin{align} \label{subnradiationkernel}
    \mathcal{R}^{(n),\mu\nu}(\bar{k}) &=  \frac{\kappa^3 m_1 m_2}{4} \int \hat{d}^4 \bar{q} \Big[ \sum_{r=0}^{n} \frac{1}{(n-r)!} e^{-i b\cdot \bar{q}} \hat{\delta}(u_1 \cdot \bar{q}) \hat{\delta}(u_2 \cdot \bar{q}) (i b \cdot \bar{k})^{n-r} K_{r-1}^{\mu\nu}\cr
    &\hspace{1cm} + \sum_{r=0}^{n-1} \frac{(-1)^{n-r}}{(n-r)!}\Big\{e^{-i b\cdot \bar{q}} \hat{\delta}^{(n-r)}(u_1 \cdot \bar{q}) \hat{\delta}(u_2 \cdot \bar{q}) (u_1 \cdot \bar{k})^{n-r}\Big( K_{r-1}^{\mu\nu}\Big) + e^{i b\cdot \bar{q}} \Big( 1 \leftrightarrow 2 \Big)\Big\}\cr
&\hspace{1cm}+ \sum_{r=0}^{n-2}\sum_{\substack{t,s \geq 1 \\
\ni (t+s) = n-r}}\frac{(-1)^s}{t! s!}e^{-i b\cdot \bar{q}} (ib \cdot \bar{k})^t (u_1 \cdot \bar{k})^s \hat{\delta}^{(s)}(u_1 \cdot \bar{q}) \hat{\delta}(u_2 \cdot \bar{q}) K_{r-1}^{\mu\nu}  
\Big] \,,
\end{align}
where the polynomial $K_{n}^{\mu\nu}$ is defined in Section \ref{sec:treesoft}. Here, the superscript in $ \mathcal{R}^{(n),\mu\nu}$ denotes the truncation of the radiation kernel to $\mathcal{O}(\omega^{n-1})$.

Using the integrals evaluated in Appendix \ref{allorderintegrals}, we will isolate the logarithmic contributions ($\omega^{n-1}\log{(\omega b)}$) that come only from the following two types of integrals:
\begin{align}
	I_1^{\mu\nu} &=\int \hat{d}^4 \bar{q} \frac{1}{(n-1)!} e^{-i b\cdot \bar{q}} \hat{\delta}(u_1 \cdot \bar{q}) \hat{\delta}(u_2 \cdot \bar{q}) (i b \cdot \bar{k})^{n-1} K_{0}^{\mu\nu} \cr
	&=-\frac{1}{(n-1)!}\int \hat{d}^4 \bar{q}  e^{-i b\cdot \bar{q}} \hat{\delta}(u_1 \cdot \bar{q}) \hat{\delta}(u_2 \cdot \bar{q}) (i b \cdot \bar{k})^{n-1} \Big[\frac{2\gamma}{ \bq^2} \Big(-  u_1^{(\mu} u_2^{\nu)} + \frac{2\omega_2}{\omega_1} u_1^\mu u_1^\nu  \Big) \cr
	&\hspace{9cm}+\frac{2}{\bq^2}\Big(\gamma^2 - \frac{1}{2} \Big)\frac{\bq^\mu \bq^\nu}{\bq^2} \Big] \cr
    &=\frac{i^{n-1}\gamma}{(n-1)!\pi \gamma\beta} (\omega b)^{n-1} \log{(\omega b)}\Big(  u_1^{(\mu} u_2^{\nu)} - \frac{(u_2 \cdot \bar{k})}{(u_1 \cdot \bar{k})} u_1^{(\mu} u_1^{\nu)}  \Big) + \mathcal{O}(\omega^{n-1}) \,,
\end{align}
where we have used the integral result of equation \eqref{i1}
and
\begin{align}
	I_2^{\mu\nu} &= -\frac{1}{(n-1)!}e^{-i b\cdot \bar{q}} (ib \cdot \bar{k})^{n-1} (u_1 \cdot \bar{k}) \hat{\delta}^{\prime}(u_1 \cdot \bar{q}) \hat{\delta}(u_2 \cdot \bar{q}) K_{-1}^{\mu\nu} \cr
	&=\frac{1}{(n-1)!} \int \hat{d}^4 \bar{q}  e^{-i b\cdot \bar{q}} \hat{\delta}^{\prime} (u_1 \cdot \bar{q}) \hat{\delta}(u_2 \cdot \bar{q}) (u_1 \cdot \bar{k}) (i b \cdot \bar{k})^{n-1} \frac{1}{\bq^2}\Big(\gamma^2 - \frac{1}{2} \Big)\Big\{- \frac{1}{\omega_1} \bq^{(\mu} u_1^{\nu)} \cr
	&\hspace{9cm}- \frac{1}{\omega_1^2}(\bar{k} \cdot \bq) u_1^\mu u_1^\nu  \Big\} \cr
    &=-\frac{i^{n-1}}{2(n-1)!\pi \gamma^3 \beta^3 } (\omega b)^{n-1} \log{(\omega b)} \Big(\gamma^2 - \frac{1}{2} \Big)\Big\{(\gamma u_2 - u_1)^{(\mu} u_1^{\nu)}  \cr
    &\hspace{7cm} - \frac{1}{(u_1 \cdot \bar{k})}(\bar{k} \cdot (\gamma u_2 - u_1)) u_1^\mu u_1^\nu  \Big\} \,,
\end{align}
where we have used the integral result of equation \eqref{derintegrals}.\\
Therefore upon simplifying, the log term in (sub)$^n$-leading radiation kernel corresponding to particle 1 is given by
\begin{align} \label{subnlogsoftexact}
    \mathcal{R}_1^{(\omega b)^{n-1}\log{(\omega b)},\mu\nu} &= \frac{i^{n-1}m_1 m_2\kappa^3}{4\pi(n-1)!\gamma^3\beta^3}\gamma (2\gamma^2 - 3) (\omega b)^{n-1} \log{(\omega b)} \Big( u_1^{(\mu} u_2^{\nu)} - \frac{(u_2 \cdot \bar{k})}{(u_1 \cdot \bar{k})} u_1^{(\mu} u_1^{\nu)}  \Big) \,.
\end{align}
 Note that in the deflection less limit ($\vert b\vert \rightarrow\, \infty$) such that $\omega b $ is fixed, the logarithmic contributions survive. The rest of the terms constitute integrals similar to the ones described in section \ref{subnsubsection} and they do not give any logarithmic contributions. Also, the terms obtained using quantum soft theorems and then taking the classical limit, match with the counterparts in the soft expansion of the radiation kernel in equation \eqref{subnradiationkernel}.

\section{Radiation kernel to (sub)$^n$-leading order in frequency} \label{subnsection}
In this section, we prove our main result. Given a (sub)$^{n}$-leading soft graviton theorem of a tree-level gravitational amplitude, the so-called remainder terms never contribute to the leading log contribution that arises in the classical limit. 

\subsection{(sub)$^n$-leading order soft radiation from quantum soft theorems} \label{subnsubsection}
We will now compute the soft radiation by applying (sub)$^n$-leading soft graviton operator on the quantum four-point amplitude and then take the classical limit. We will see that not all the terms in the soft expansion of the radiation kernel can be recovered by applying the soft theorems and the remainder terms do not correspond to any logarithmic contribution. As $\omega^{n-1}\log{\omega}$ is more dominant than the $\omega^{n-1}$ terms, the low-frequency classical radiation during a scattering process is simply obtained from the soft theorems. One can discard the remainder terms then.\\
From quantum soft theorems, the (sub)$^n$-leading radiation kernel is given by
\begin{align} \label{subnleadingsoft}
    \mathcal{R}^{\mu\nu}_{\omega^{(n-1)}} &= \frac{1}{4 m_1 m_2}\int \hat{d}^{4} q_1 \hat{d}^4 q_2 e^{i q_1 \cdot b} \hat{\delta}(u_1 \cdot q_1) \hat{\delta}(u_2 \cdot q_2) \cr
    &\hspace{3cm} \times \frac{\kappa}{2} \sum_{i=1,2} \Big[\frac{J_i^{\mu\rho}k_\rho J_i^{\nu\sigma} k_\sigma}{p_i \cdot k}\Big(k \cdot \frac{\partial}{\partial p_i} \Big)^{n-2} + \frac{\tilde{J}_i^{\mu\rho}k_\rho \tilde{J}_i^{\nu\sigma} k_\sigma}{\tilde{p}_i \cdot k}\Big(k \cdot \frac{\partial}{\partial \tilde{p}_i} \Big)^{n-2}  \Big] \cr
    &\hspace{9cm}\Big(\hat{\delta}^{(4)}(q_1 + q_2) \mathcal{A}_4 \Big) \,,
\end{align}
where
\begin{align}
    \mathcal{A}_4 [p_1, \tilde{p}_1, p_2, \tilde{p}_2] &= \frac{\kappa^2}{2q_2^2} \Big[(p_2 \cdot \tilde{p}_2)(m_1^2 - p_1 \cdot \tilde{p}_1) + m_2^2 (p_1 \cdot \tilde{p}_1 - 2 m_1^2) \cr
    &\hspace{4cm} + (p_1 \cdot \tilde{p}_2)(p_2 \cdot \tilde{p}_1) + (p_1 \cdot p_2)(\tilde{p}_1 \cdot \tilde{p}_2)\Big] \,.
\end{align}
From now on, we will consider the contribution from particle 1. First, let us evaluate the soft operators' action on $\mathcal{A}_4$. The classical contribution to the soft radiation in this case comes from the action of the soft operators on the denominator of the amplitude and is given by  
\begin{align} 
     \mathcal{R}^{\mu\nu}_{\omega^{(n-1)},A} &= (-1)^{n+1} \frac{ \kappa^3 m_1 m_2}{4} \int \hat{d}^{4} \bq e^{-i \bq \cdot b} \hat{\delta}(u_1 \cdot \bq) \hat{\delta}(u_2 \cdot \bq)  \frac{2^{n-1}}{(\bq^2)^{n+1}} (\bq \cdot \bar{k})^{n-1} \cr &\hspace{5cm} \times \left(\bq^\mu \bq^\nu \Big(\gamma^2 -\frac{1}{2}\Big) + \bq^2 \frac{(u_2 \cdot \bar{k})}{(\bq \cdot \bar{k})}\bq^{(\mu} u_1^{\nu)}\right)  \,.
\end{align}
This term doesn't give any logarithmic contributions using the integral results of Appendix \ref{allorderintegrals}.
Let us evaluate the soft operators' action on the delta function now where we use the following distributional identity:
\begin{align} \label{distrib_identity}
    S^{(n),\mu\nu}\hat{\delta}^{(4)}(q_1 + q_2) &= \hat{\delta}^{(4)}(q_1 + q_2) S^{(n),\mu\nu} - (k \cdot \partial)\hat{\delta}^{(4)}(q_1 + q_2) S^{(n-1),\mu\nu} \cr
    &+ \frac{1}{2}(k \cdot \partial)^2 \hat{\delta}^{(4)}(q_1 + q_2) S^{(n-2),\mu\nu} + \cdots + \frac{(-1)^n}{n!}(k \cdot \partial)^n \hat{\delta}^{(4)}(q_1 + q_2) S^{(0),\mu\nu}\,. \cr
\end{align}
The soft radiation in this case is given by
\begin{align}
     \mathcal{R}^{\mu\nu}_{\omega^{(n-1)},D} &= \frac{1}{m_1 m_2}\int \hat{d}^{4} q_1 \hat{d}^4 q_2 e^{i q_1 \cdot b} \hat{\delta}(u_1 \cdot q_1) \hat{\delta}(u_2 \cdot q_2) \Big[-(k \cdot \partial)\hat{\delta}^{(4)}(q_1 + q_2) S^{(n-1),\mu\nu} \cr
     &+ \frac{1}{2}(k \cdot \partial)^2 \hat{\delta}^{(4)}(q_1 + q_2) S^{(n-2),\mu\nu} + \cdots +
     \frac{(-1)^{n-2}}{(n-2)!}(k \cdot \partial)^{n-2}\hat{\delta}^{(4)}(q_1 + q_2) S^{(2),\mu\nu} \cr
     &+ \frac{(-1)^{n-1}}{(n-1)!}(k \cdot \partial)^{n-1}\hat{\delta}^{(4)}(q_1 + q_2) S^{(1),\mu\nu} + \frac{(-1)^{n}}{n!}(k \cdot \partial)^n \hat{\delta}^{(4)}(q_1 + q_2) S^{(0),\mu\nu} \Big] \mathcal{A}_4 \,. \cr
\end{align}
We give a detailed derivation of the computation in Appendix \ref{subnsection_detailed}. Here we simply isolate the logarithmic contributions to the classical soft radiation which comes from the following two terms:

\begin{align}
        \mathcal{R}^{\mu\nu}_{\omega^{(n-1)},4} &= \frac{1}{m_1 m_2}\int \hat{d}^{4} q_1 \hat{d}^4 q_2 e^{i q_1 \cdot b} \hat{\delta}(u_1 \cdot q_1) \hat{\delta}(u_2 \cdot q_2) \frac{(-1)^{n-1}(k \cdot \partial)^{n-1}}{(n-1)!}\hat{\delta}^{(4)}(q_1 + q_2) S^{(1),\mu\nu} \mathcal{A}_4 \,. 
        \end{align}
Integrating $q_1$  and relabelling $q_2 \rightarrow q$ and keeping only $\mathcal{O}(\omega^{n-1})$ terms, we have, 
\begin{align}
    \mathcal{R}^{\mu\nu}_{\omega^{(n-1)},4} &= \frac{1}{m_1 m_2} \int \hat{d}^4 q e^{-i q\cdot b}\hat{\delta}(u_2 \cdot q) \Big\{\frac{(i k\cdot b)^{n-1}}{(n-1)!}\hat{\delta}(u_1 \cdot q)  \cr
    &\hspace{3cm}+\sum_{\substack{r,s \geq 1\\
\ni (r+s) = n-1}}\frac{(-1)^s}{r! s!} (ib \cdot k)^r (u_1 \cdot k)^s \hat{\delta}^{(s)}(u_1 \cdot q) \cr
    &\hspace{3cm} + \frac{(-1)^{n-1}}{(n-1)!} (u_1 \cdot k)^{n-1} \hat{\delta}^{(n-1)}(u_1 \cdot q)\Big\} S^{(1),\mu\nu} \mathcal{A}_4 \cr
    &=\frac{i^{n-1} m_1 m_2 \kappa^3 \gamma}{(n-1)!\pi\gamma\beta} (\omega b)^{n-1} \log{(\omega b)} u_1^{(\mu} \Big(u_2^{\nu)}  - u_1^{\nu)} \frac{(\bar{k} \cdot u_2)}{(\bar{k} \cdot u_1)} \Big) + \mathcal{O}(\omega^{n-1})\,.
\end{align}
where we have used the integral result of equation \eqref{i1},
and
\begin{align}
    \mathcal{R}^{\mu\nu}_{\omega^{(n-1)},5} &= \frac{1}{m_1 m_2} \int \hat{d}^{4} q_1 \hat{d}^4 q_2 e^{i q_1 \cdot b} \hat{\delta}(u_1 \cdot q_1) \hat{\delta}(u_2 \cdot q_2) \frac{(-1)^{n}(k \cdot \partial)^{n}}{n!}\hat{\delta}^{(4)}(q_1 + q_2) S^{(0),\mu\nu} \mathcal{A}_4 \,.
\end{align}
Integrating $q_1$  and relabelling $q_2 \rightarrow q$ and keeping only $\mathcal{O}(\omega^{n-1})$ terms, we have, 
\begin{align}
    \mathcal{R}^{\mu\nu}_{\omega^{(n-1)},5} &= \frac{1}{m_1 m_2} \int \hat{d}^4 q e^{-i q\cdot b}\hat{\delta}(u_2 \cdot q) \Big\{\frac{(i k\cdot b)^{n}}{n!}\hat{\delta}(u_1 \cdot q)  \cr
    &+\sum_{\substack{r,s \\
\ni (r+s) = n}}\frac{(-1)^s}{r! s!} (ib \cdot k)^r (u_1 \cdot k)^s \hat{\delta}^{(s)}(u_1 \cdot q) + \frac{(-1)^{n}}{n!} (u_1 \cdot k)^{n} \hat{\delta}^{(n)}(u_1 \cdot q)\Big\} S^{(0),\mu\nu} \mathcal{A}_4 \cr
&=- \kappa^3(2(p_1 \cdot p_2)^2 - m_1^2 m_2^2)\frac{i^{n-1}}{4\pi(n-1)!\gamma^3 \beta^3} (\omega b)^{n-1}\log{(\omega b)} \cr
     &\hspace{2cm} \times \Big((\gamma u_2 - u_1)^{(\mu}u_1^{\nu)}- \frac{((\gamma u_2 -u_1) \cdot \bar{k})u_1^{(\mu} u_1^{\nu)}}{(u_1 \cdot \bar{k})} \Big) + \mathcal{O}(\omega^{n-1})\,,
 \end{align}
where we have used the integral result of equation \eqref{derintegrals}. The detailed derivations of the above steps are given in Appendix \ref{subnsection_detailed}.
We collect the log terms and upon simplifying, the $\omega^{n-1} \log{\omega}$ terms of radiation kernel w.r.t particle 1 from the quantum soft theorems is given by
\begin{align}
    \mathcal{R}^{\mu\nu}_{\omega^{n-1}\log{\omega}} &= \frac{i^{n-1}m_1 m_2\kappa^3}{4\pi(n-1)!\gamma^3\beta^3}\gamma (2\gamma^2 - 3) (\omega b)^{n-1} \log{(\omega b)} \Big( u_1^{(\mu} u_2^{\nu)} - \frac{(u_2 \cdot \bar{k})}{(u_1 \cdot \bar{k})} u_1^{(\mu} u_1^{\nu)}  \Big) \,,
\end{align}
which matches with the log terms of (sub)$^n$-leading order soft expansion of the radiation kernel in equation \eqref{subnlogsoftexact}. One can also see that the rest of the terms obtained using quantum soft theorems also match with the counterparts in the soft expansion of the radiation kernel in equation \eqref{subnradiationkernel}.

\subsection{Remainder terms in (sub)$^n$-leading order soft radiation}
Comparing the soft expansion of the radiation kernel and the radiation kernel obtained using quantum soft theorems to (sub)$^n$-leading order in frequency, we see that not all the terms in the soft expansion of the radiation kernel in equation \eqref{subnradiationkernel} are recovered by applying the soft theorems. Such terms in the (unstripped) five-point amplitude do not factorize as
soft factors times the four-point amplitude. These are known as the ``Remainder terms.'' We have identified such terms at (sub)$^n$-leading order for $ n \geq 3$ in the soft radiation kernel given by
\begin{align}
    \mathcal{X}^{\mu\nu}_{\mathcal{R},\omega^{(n-1)}} &= \frac{\kappa^3 m_1 m_2}{4} \int \hat{d}^4 \bar{q} \Big[ \sum_{r=3}^{n} \frac{1}{(n-r)!} e^{-i b\cdot \bar{q}} \hat{\delta}(u_1 \cdot \bar{q}) \hat{\delta}(u_2 \cdot \bar{q}) (i b \cdot \bar{k})^{n-r} \Lambda_{r-1}^{\mu\nu}\cr
    &\hspace{1cm} + \sum_{r=3}^{n-1} \frac{(-1)^{n-r}}{(n-r)!}\Big\{e^{-i b\cdot \bar{q}} \hat{\delta}^{(n-r)}(u_1 \cdot \bar{q}) \hat{\delta}(u_2 \cdot \bar{q}) (u_1 \cdot \bar{k})^{n-r}\Big( \Lambda_{r-1}^{\mu\nu}\Big) + e^{i b\cdot \bar{q}} \Big( 1 \leftrightarrow 2 \Big)\Big\}\cr
&\hspace{1cm}+ \sum_{r=3}^{n-2}\sum_{\substack{t,s \geq 1 \\
\ni (t+s) = n-r}}\frac{(-1)^s}{t! s!}e^{-i b\cdot \bar{q}} (ib \cdot \bar{k})^t (u_1 \cdot \bar{k})^s \hat{\delta}^{(s)}(u_1 \cdot \bar{q}) \hat{\delta}(u_2 \cdot \bar{q}) \Lambda_{r-1}^{\mu\nu}  
\Big] \,, \cr
\end{align}
where the polynomial $\Lambda_{n}^{\mu\nu}$ is defined in Section \ref{sec:treesoft}.
Using the integral results of Appendix \ref{allorderintegrals}, it is evident that these remainder terms do not give any logarithmic contributions. As $\omega^{n-1}\log{\omega}$ is more dominant than the $\omega^{n-1}$ terms, one can simply discard the remainder terms in computing the low-frequency classical radiation during a scattering process.

\section{Radiation kernel to (sub)$^3$-leading order in frequency} \label{sub3section}
In this section, we will compute the soft radiative gravitational field to (sub)$^3$-leading order in frequency. One can simply substitute $n=3$ in the previous section for the analysis. We will only summarise the important results here. 
\begin{itemize}
    \item The leading logarithmic contribution to the radiation kernel in this order is given by
        \begin{align} 
    \mathcal{R}_{\omega^2\log{\omega}}^{\mu\nu} &= -\frac{\kappa^3 m_1 m_2}{8\pi \gamma^3 \beta^3} (\omega b)^2 \log{ (\omega b)}  \gamma (2\gamma^2 - 3) \Big( u_1^{(\mu} u_2^{\nu)} - \frac{(u_2 \cdot \bar{k})}{(u_1 \cdot \bar{k})} u_1^{(\mu} u_1^{\nu)}  \Big) + (1 \leftrightarrow 2)\,.
\end{align}
As stated before, in the deflection less limit ($\vert b\vert \rightarrow\, \infty$) such that $\omega b $ is fixed, the logarithmic contribution survives.
    \item The factorized terms in the radiation kernel that are obtained via the quantum soft graviton theorems match with the counterparts obtained from the soft expansion of the classical radiation kernel at (sub)$^3$-leading order. However, at this order $(n \geq 3)$, the radiation kernel is infected with the presence of non-factorizing remainder terms.
    \item We have identified such remainder terms at (sub)$^3$-leading order in the soft radiation kernel given by
\begin{align}
    \mathcal{X}^{\mu\nu}_{\mathcal{R},\omega^2} &=\frac{\kappa^3 m_1 m_2}{4} \int \hat{d}^4 \bar{q} \Big\{ e^{-i b\cdot \bar{q}} \hat{\delta}(u_1 \cdot \bar{q}) \hat{\delta}(u_2 \cdot \bar{q}) H_2^{\mu\nu} + e^{i b\cdot \bar{q}} \Big( 1 \leftrightarrow 2 \Big)\Big\} \,,
\end{align}
where
\begin{align}
    H_2^{\mu\nu} = -\frac{4}{(\bq^2)^2} \Big( \omega_2^2 u_1^\mu u_1^\nu - \frac{\omega_1 \omega_2}{2}(u_2^\mu u_1^\nu + u_2^\nu u_1^\mu)\Big) \,.
\end{align}
As expected, the remainder term does not give any logarithmic contributions. As $\omega^2\log{\omega}$ is more dominant than the $\omega^2$ terms, one can simply discard the remainder term in computing the low-frequency classical radiation during a scattering process.
\end{itemize}

\section{Discussion} \label{discussion}
In this work, we have shown that the tree-level (sub)$^{n}$-leading soft graviton theorems for two massive scalar fields minimally coupled to gravity generate all the logarithmic terms in the soft expansion in the limit of vanishing deflection. It would be interesting to explore the effect of the non-universal terms in the soft factors which are generated by irrelevant terms in the Lagrangian. Already at the (sub)$^{2}$-leading order in the soft expansion, where the remainder terms are zero, the corresponding soft factor is modified by the presence of a finite set of higher derivative terms in the Lagrangian \cite{Elvang:2016qvq}. Higher derivative terms will certainly change the higher-order tree-level soft factors, but we believe that they will not alter the leading logs in the deflection-less limit. However, this needs to be investigated further.

In detail, we have shown that in the deflection less limit ($\vert b\vert \rightarrow\, \infty$) such that $\omega |b| $ is fixed, all the log terms of the form $(\omega b)^{n}\log{(\omega b)}$ survive and can be completely determined by the (sub)$^{n}$-leading soft graviton theorems for tree-level gravitational amplitudes. The source of such radiative modes is the asymptotic interaction between the incoming or outgoing states, leading to the emission of gravitational radiation only from $t\, \rightarrow\, \pm \infty$. However, there could be loop corrections to the result and would need to be investigated further. The universal log terms of the form $(\omega b)^n \log{(\omega b)}^{n+1}$, $n \geq 1$ arise from the higher-loop amplitudes which survive in the $\omega \rightarrow 0, |b| \rightarrow \infty$ limit such that $\omega b$ is fixed. The loop computations would be significantly simpler as we are interested only in the infinite impact parameter limit where $p_i^\prime = p_i$. Here $p_i$ and $p_i^\prime$ are the initial and final momenta of the particles.

It would be interesting to compute the soft gravitational radiation for $D >4$ and analyze the soft spectra. In contrast to the case in $D=4$, the sub-leading contribution in higher dimensions arises from the integration region where $|q| \sim b^{-1}$. This aligns with the classical soft theorem in higher dimensions, as discussed in \cite{Laddha:2019yaj}, where it was shown that during scattering, the ``outer'' space-time region with size $\geq b$ contributes to the sub-leading radiation. There is a reversal of order in the behavior of soft emission between $D = 4$ and $D > 4$. In dimensions higher than four, the integral yields $\omega^0$ terms from the ``UV region'' where $|q| \sim b^{-1}$, and terms proportional to $\omega^{D-4}$ from the ``IR region'' where $|l| \geq \omega$. In $D=4$, however, the integral produces logarithmic terms $\log{\omega}$ from the IR region and $\omega^0$ terms from the UV region. It would be worth examining whether any logarithmic contributions appear at (sub)$^n$-leading order in $D=5$, as this represents the first non-trivial case, and also reviewing the remainder terms. This would prove to be highly useful to have an interpretation of the classical soft theorems in $D>4$ spacetime dimensions.
\acknowledgments
I am grateful to Alok Laddha for suggesting the problem, numerous insightful discussions, and constant encouragement. I thank him for going through the draft and providing valuable suggestions and comments on it. I thank Sujay K. Ashok for constant encouragement and valuable comments on the draft. I would also like to thank Arkajyoti Manna and Akavoor Manu for useful discussions.  

\begin{appendix}
\section{Conventions}\label{convention}
Throughout the paper, we will use the metric signature as $(+,-,-,-)$, unless otherwise stated. So, the on-shell condition is $p^2=m^2$. Since the impact parameter is spacelike we have $-b^2 > 0$. 
The rescaled delta functions appearing in the main text are defined as
\begin{equation}
	\hat{\delta}(p\cdot q) := 2\pi\delta(p\cdot q),\ \ \ \ \ \hat{\delta}^{(4)}(p + q) := (2\pi)^{4}\delta^{(4)}(p + q).
\end{equation}
where $p^{\mu}$ and $q^{\mu}$ are generic four vectors. We also absorb the $2\pi$ factor in the measure $d^4q$ and define the rescaled measure as
\begin{equation}
	\hat{d}^{4}q := \frac{d^{4}q}{(2\pi)^{4}}.
\end{equation}

\section{(sub)$^2$-leading order soft radiation from quantum soft theorems} \label{sec:subsubsoft}
In this appendix, we will review the computation of the soft radiation by applying (sub)$^2$-leading soft graviton operator on the quantum four-point amplitude and then take the classical limit.\\
From quantum soft theorems, the (sub)$^2$-leading radiation kernel is given by
\begin{align} \label{subsubleadingsoft}
    \mathcal{R}^{\mu\nu}_\omega &= \frac{1}{4 m_1 m_2}\int \hat{d}^{4} q_1 \hat{d}^4 q_2 e^{i q_1 \cdot b/\hbar} \hat{\delta}(u_1 \cdot q_1) \hat{\delta}(u_2 \cdot q_2) \ \kappa \sum_{i=1,2} \Big[\frac{J_i^{\mu\rho}k_\rho J_i^{\nu\sigma} k_\sigma}{p_i \cdot k} + \frac{\tilde{J}_i^{\mu\rho}k_\rho \tilde{J}_i^{\nu\sigma} k_\sigma}{\tilde{p}_i \cdot k} \Big] \cr
    &\hspace{10cm}\Big(\hat{\delta}^{(4)}(q_1 + q_2) \mathcal{A}_4 \Big) \,,
\end{align}
where
\begin{align}
    \mathcal{A}_4 [p_1, \tilde{p}_1, p_2, \tilde{p}_2] &= \frac{\kappa^2}{2q_2^2} \Big[(p_2 \cdot \tilde{p}_2)(m_1^2 - p_1 \cdot \tilde{p}_1) + m_2^2 (p_1 \cdot \tilde{p}_1 - 2 m_1^2) \cr
    &\hspace{4cm}+ (p_1 \cdot \tilde{p}_2)(p_2 \cdot \tilde{p}_1) + (p_1 \cdot p_2)(\tilde{p}_1 \cdot \tilde{p}_2)\Big] \,.
\end{align}
First, let us evaluate the soft operators' action on $\mathcal{A}_4$. We consider the contribution from particle 1 for now. The action of the soft operators on the numerator of the amplitudes is given by
\begin{align}
    \kappa \frac{J_1^{\mu\rho}k_\rho J_1^{\nu\sigma} k_\sigma}{p_1 \cdot k}  [\mathcal{A}_4]_N &= -\kappa^3  \frac{k_\rho k_\sigma}{2q_2^2 (p_1 \cdot k)} \Big(p_1 \wedge \frac{\partial}{\partial p_1} \Big)^{\mu\rho} \Big(p_1 \wedge \frac{\partial}{\partial p_1} \Big)^{\nu\sigma} \Big[(p_2 \cdot \tilde{p}_2)(m_1^2 - p_1 \cdot \tilde{p}_1) \cr
    &\hspace{2cm}+ m_2^2 (p_1 \cdot \tilde{p}_1 - 2 m_1^2) + (p_1 \cdot \tilde{p}_2)(p_2 \cdot \tilde{p}_1) + (p_1 \cdot p_2)(\tilde{p}_1 \cdot \tilde{p}_2)\Big]\cr
    &=0 \,.
\end{align}
and
\begin{align}
    \kappa \frac{\tilde{J}_1^{\mu\rho}k_\rho \tilde{J}_1^{\nu\sigma} k_\sigma}{\tilde{p}_1 \cdot k}  [\mathcal{A}_4]_N &= -\kappa^3 \frac{k_\rho k_\sigma}{2 q_2^2 (\tilde{p}_1 \cdot k)} \Big(\tilde{p}_1 \wedge \frac{\partial}{\partial \tilde{p}_1} \Big)^{\mu\rho} \Big(\tilde{p}_1 \wedge \frac{\partial}{\partial \tilde{p}_1} \Big)^{\nu\sigma} \Big[(p_2 \cdot \tilde{p}_2)(m_1^2 - p_1 \cdot \tilde{p}_1) \cr
    &\hspace{2cm}+ m_2^2 (p_1 \cdot \tilde{p}_1 - 2 m_1^2) + (p_1 \cdot \tilde{p}_2)(p_2 \cdot \tilde{p}_1) + (p_1 \cdot p_2)(\tilde{p}_1 \cdot \tilde{p}_2)\Big] \cr
    &=0 \,.
\end{align} 
The classical contribution from the action of the soft operators on the denominator of the amplitude is given by
\begin{align} \label{subsubclassicalamp}
    -\frac{\kappa^3}{2(\bq^2)^3} (\bq \cdot \bar{k}) \left(\bq^\mu \bq^\nu ((p_1 \cdot p_2)^2 - \frac{1}{2} m_1^2 m_2^2) + \bq^2 \frac{(p_2 \cdot \bar{k})}{(\bq \cdot \bar{k})}\bq^{(\mu} p_1^{\nu)}\right) \,.
\end{align}
Therefore the classical contribution to soft radiation from the action of the (sub)$^2$-leading soft operator ($S^{(2),\mu\nu}$) on the four-point amplitude alone is given by
\begin{align}
	\mathcal{R}_{\omega,A}^{\mu\nu} = &= -\frac{ \kappa^3 m_1 m_2}{4} \int \hat{d}^{4} \bq e^{-i \bq \cdot b} \hat{\delta}(u_1 \cdot \bq) \hat{\delta}(u_2 \cdot \bq)  \frac{1}{2(\bq^2)^3} (\bq \cdot \bar{k}) \cr
    &\hspace{5cm} \times \left(\bq^\mu \bq^\nu \Big(\gamma^2 -\frac{1}{2}\Big) + \bq^2 \frac{(u_2 \cdot \bar{k})}{(\bq \cdot \bar{k})}\bq^{(\mu} u_1^{\nu)}\right)\,.
\end{align} 
Let us evaluate the soft operators' action on the delta function now. Again we restrict to the contribution from particle 1. We use the distributional identity:
\begin{align}
    S^{(2),\mu\nu}\hat{\delta}^{(4)}(q_1 + q_2) &= \hat{\delta}^{(4)}(q_1 + q_2) S^{(2),\mu\nu} - (k \cdot \partial)\hat{\delta}^{(4)}(q_1 + q_2) S^{(1),\mu\nu} \cr
    &\hspace{5cm}+ \frac{1}{2}(k \cdot \partial)^2 \hat{\delta}^{(4)}(q_1 + q_2) S^{(0),\mu\nu}\,.
\end{align} Here, $S^{(0),\mu\nu}, S^{(1),\mu\nu}, S^{(2),\mu\nu}$ are the leading, sub-leading and (sub)$^2$-leading soft operators respectively. We have,
\begin{align}
     \mathcal{R}^{\mu\nu}_{\omega,D} &= \frac{1}{m_1 m_2}\int \hat{d}^{4} q_1 \hat{d}^4 q_2 e^{i q_1 \cdot b/\hbar} \hat{\delta}(u_1 \cdot q_1) \hat{\delta}(u_2 \cdot q_2) \Big[\hat{\delta}^{(4)}(q_1 + q_2) S^{(2),\mu\nu} \cr
     &\hspace{2cm}- (k \cdot \partial)\hat{\delta}^{(4)}(q_1 + q_2) S^{(1),\mu\nu} + \frac{1}{2}(k \cdot \partial)^2 \hat{\delta}^{(4)}(q_1 + q_2) S^{(0),\mu\nu} \Big] \mathcal{A}_4 \,.
\end{align}
From equation \eqref{subsubclassicalamp}, we have the classical contribution of $S^{(2),\mu\nu}$ on the amplitude. Therefore, we compute the remaining two terms. We have
\begin{align}
    \mathcal{R}^{\mu\nu}_{\omega,1} &= -\frac{1}{m_1 m_2}\int \hat{d}^{4} q_1 \hat{d}^4 q_2 e^{i q_1 \cdot b/\hbar} \hat{\delta}(u_1 \cdot q_1) \hat{\delta}(u_2 \cdot q_2) (k \cdot \partial)\hat{\delta}^{(4)}(q_1 + q_2) S^{(1),\mu\nu} \mathcal{A}_4 \cr
    &= \frac{1}{m_1 m_2}\int \hat{d}^{4} q_1 \hat{d}^4 q_2 e^{i q_1 \cdot b/\hbar} \hat{\delta}(u_1 \cdot q_1) \hat{\delta}(u_2 \cdot q_2) \Big\{\hat{\delta}^{(4)}(q_1 + q_2- k) \cr
    &\hspace{8cm}- \hat{\delta}^{(4)}(q_1 + q_2)\Big\} S^{(1),\mu\nu} \mathcal{A}_4 \,. \cr
\end{align}
Integrating $q_1$  and relabelling $q_2 \rightarrow q$, we have
\begin{align}
    \mathcal{R}^{\mu\nu}_{\omega,1} &= \int \hat{d}^4 q e^{-i q\cdot b/\hbar}\hat{\delta}(u_2 \cdot q) \Big\{\hat{\delta}(u_1 \cdot (k -q))e^{i k \cdot b/\hbar} - \hat{\delta}(u_1 \cdot q)\Big\} S^{(1),\mu\nu} \mathcal{A}_4 \,.
\end{align}
Writing only the $\mathcal{O}(\omega)$ term,
\begin{align}
    \mathcal{R}^{\mu\nu}_{\omega,1} &= \int \hat{d}^4 q e^{-i q\cdot b/\hbar}\hat{\delta}(u_2 \cdot q) \Big\{\hat{\delta}(u_1 \cdot q) (i k\cdot \frac{b}{\hbar}) - (u_1 \cdot k) \hat{\delta}^\prime (u_1 \cdot q)\Big\} S^{(1),\mu\nu} \mathcal{A}_4 \,.
\end{align}
The sub-leading soft graviton operator is given by
\begin{align} \label{subsoft}
    S^{(1),\mu\nu} \mathcal{A}_4 &= \frac{\kappa}{2} \Big[\frac{p_1^{(\mu}J_1^{\nu)\rho}k_\rho}{(p_1 \cdot k)} - \frac{\tilde{p}_1^{(\mu}\tilde{J}_1^{\nu)\rho}k_\rho}{(\tilde{p}_1 \cdot k)}\Big]\mathcal{A}_4 \,.
\end{align}
The action of the soft operators on the amplitude is given by
\begin{align}
    \kappa \frac{p_1^{(\mu}J_1^{\nu)\rho}k_\rho}{(p_1 \cdot k)}\mathcal{A}_4 &= \frac{i\kappa}{p_1 \cdot k}\Big( p_1^{(\mu} p_1^{\nu)} (k \cdot \frac{\partial}{\partial p_1}) - (p_1 \cdot k)p_1^{(\mu}\frac{\partial}{\partial p_{1\nu)}}\Big)\mathcal{A}_4 \cr
    &= -\frac{i\kappa^3}{2q^2 (k \cdot p_1)} p_1^{\mu} \Big[(k \cdot p_1)(\tilde{p}_1^\nu (m_2^2 - p_2 \cdot \tilde{p}_2) + p_2^\nu (\tilde{p}_1 \cdot \tilde{p}_2)) \cr
    &\hspace{3cm}- p_1^\nu ((k \cdot \tilde{p}_1)(m_2^2 - p_2 \cdot \tilde{p}_2)+ (k \cdot \tilde{p}_2)(\tilde{p}_1 \cdot p_2) \cr
    &\hspace{3cm}+ (k \cdot p_2)(\tilde{p}_1 \cdot \tilde{p}_2)) + \tilde{p}_2^\nu (k \cdot p_1)(\tilde{p}_1 \cdot p_2) \cr
    &- \frac{1}{2 q^2} \Big(p_1^\nu (k \cdot \tilde{p}_1 + k \cdot p_2 - k \cdot \tilde{p}_2) -(p_1 \cdot k)(\tilde{p}_1^\nu + p_2^\nu - \tilde{p}_2^\nu) \Big) (2(p_1 \cdot p_2)^2 - m_1^2 m_2^2) \Big] \,. \cr
\end{align}
The classical contribution is given by
\begin{align} \label{subsoft1}
    \kappa \frac{p_1^{(\mu}J_1^{\nu)\rho}k_\rho}{(p_1 \cdot k)}\mathcal{A}_4 &= -\frac{i\kappa^3}{\bq^2 (\bar{k} \cdot p_1)} p_1^{(\mu} \Big[p_2^{\nu)} (\bar{k} \cdot p_1)(p_1 \cdot p_2) - p_1^{\nu)} (\bar{k} \cdot p_2)(p_1 \cdot p_2) \Big] \,.
\end{align}
and
\begin{align}
    \kappa \frac{\tilde{p}_1^{(\mu}\tilde{J}_1^{\nu)\rho}k_\rho}{(\tilde{p}_1 \cdot k)}\mathcal{A}_4 &= \frac{-i\kappa}{\tilde{p}_1 \cdot k}\Big( \tilde{p}_1^{(\mu} \tilde{p}_1^{\nu)} (k \cdot \frac{\partial}{\partial \tilde{p}_1}) - (\tilde{p}_1 \cdot k)\tilde{p}_1^{(\mu}\frac{\partial}{\partial \tilde{p}_{1\nu)}}\Big)\mathcal{A}_4 \cr
    &= \frac{i\kappa^3}{2q^2 (k \cdot \tilde{p}_1)}\tilde{p}_1^{\mu} \Big[(k \cdot \tilde{p}_1)(p_1^\nu (m_2^2 - p_2 \cdot \tilde{p}_2) + p_2^\nu (p_1 \cdot \tilde{p}_2)) \cr
    &\hspace{3cm}- \tilde{p}_1^\nu ((k \cdot p_1)(m_2^2 - p_2 \cdot \tilde{p}_2)+ (k \cdot \tilde{p}_2)(p_1 \cdot p_2) \cr
    &\hspace{3cm}+ (k \cdot p_2)(p_1 \cdot \tilde{p}_2)) + \tilde{p}_2^\nu (k \cdot \tilde{p}_1)(p_1 \cdot p_2) \cr
    &- \frac{1}{2 q^2} \Big(\tilde{p}_1^\nu (k \cdot p_1 + k \cdot \tilde{p}_2 - k \cdot p_2) -(\tilde{p}_1 \cdot k)(p_1^\nu + \tilde{p}_2^\nu - p_2^\nu) \Big) (2(p_1 \cdot p_2)^2 - m_1^2 m_2^2) \Big] \,. \cr
\end{align}
The classical contribution is given by
\begin{align} \label{subsoft2}
    \kappa \frac{\tilde{p}_1^{(\mu}\tilde{J}_1^{\nu)\rho}k_\rho}{(\tilde{p}_1 \cdot k)}\mathcal{A}_4 &= \frac{i\kappa^3}{\bq^2 (\bar{k} \cdot p_1)}  \Big[p_1^{(\mu} p_2^{\nu)} (\bar{k} \cdot p_1)(p_1 \cdot p_2) -p_1^{(\mu} p_1^{\nu)} (\bar{k} \cdot p_2)(p_1 \cdot p_2) \cr 
    &\hspace{4cm}-\frac{1}{2 \bq^2} \bq^\mu \bq^\nu (\bar{k} \cdot p_1) (2(p_1 \cdot p_2)^2 - m_1^2 m_2^2) \Big] \,. 
\end{align}
Therefore, substituting equations \eqref{subsoft1} and \eqref{subsoft2} in equation \eqref{subsoft} the classical contribution of $\mathcal{R}^{\mu\nu}_{\omega,1}$ is given by
\begin{align} \label{subsoftclasical}
    \mathcal{R}^{\mu\nu}_{\omega,1} &= \frac{1}{m_1 m_2}\int \frac{\hat{d}^4 \bq}{\bq^2} e^{-i \bq\cdot b}\hat{\delta}(u_2 \cdot \bq) \Big\{\hat{\delta}(u_1 \cdot \bq) (i \bar{k}\cdot b) - (u_1 \cdot \bar{k}) \hat{\delta}^\prime (u_1 \cdot \bq)\Big\} \cr
    &\hspace{2cm}\times  -\frac{2i\kappa^3}{(\bar{k} \cdot p_1)} \Big[p_1^{(\mu} p_2^{\nu)} (\bar{k} \cdot p_1)(p_1 \cdot p_2) - p_1^{(\mu} p_1^{\nu)} (\bar{k} \cdot p_2)(p_1 \cdot p_2) + \frac{1}{2\bq^2} \bq^\mu \bq^\nu (\bar{k} \cdot p_1)\Big] \cr
    &= \frac{\kappa^3 m_1 m_2 \gamma}{\pi \gamma\beta } u_1^{(\mu} \Big[u_2^{\nu)} - u_1^{\nu)} \frac{(\bar{k} \cdot u_2)}{(\bar{k} \cdot u_1)}\Big] \omega b \log{(\omega b)}  + \mathcal{O}(\omega) \,,
\end{align}
where we have used the integral result of equation \eqref{i1}.
We are now left with computing one last term.
\begin{align}
    \mathcal{R}^{\mu\nu}_{\omega,2} &= \frac{1}{2 m_1 m_2}\int \hat{d}^{4} q_1 \hat{d}^4 q_2 e^{i q_1 \cdot b} \hat{\delta}(u_1 \cdot q_1) \hat{\delta}(u_2 \cdot q_2) (k \cdot \partial)^2 \hat{\delta}^{(4)}(q_1 + q_2) S^{(0),\mu\nu} \mathcal{A}_4 \cr
    &=\frac{1}{m_1 m_2}\int \hat{d}^{4} q_1 \hat{d}^4 q_2 e^{i q_1 \cdot b/\hbar} \hat{\delta}(u_1 \cdot q_1) \hat{\delta}(u_2 \cdot q_2) \Big\{\hat{\delta}^{(4)}(q_1 + q_2- k) - \hat{\delta}^{(4)}(q_1 + q_2) \cr
    &\hspace{7cm}+ (k \cdot \partial)\hat{\delta}^{(4)}(q_1 + q_2)\Big\} S^{(0),\mu\nu} \mathcal{A}_4 \,. 
\end{align}
Integrating $q_1$  and relabelling $q_2 \rightarrow q$, we have
\begin{align}
    \mathcal{R}^{\mu\nu}_{\omega,2} &= \frac{1}{m_1 m_2}\int \hat{d}^4 q e^{-i q\cdot b/\hbar}\hat{\delta}(u_2 \cdot q) \Big\{\hat{\delta}(u_1 \cdot (k -q))e^{i k \cdot b/\hbar} - \hat{\delta}(u_1 \cdot q) \cr
    &\hspace{5cm}+ \hat{\delta}(u_1 \cdot q) - \hat{\delta}(u_1 \cdot (k -q))e^{i k \cdot b/\hbar} \Big\} S^{(0),\mu\nu} \mathcal{A}_4 \,, \cr
\end{align}
where the $(k \cdot \partial)$ term is written from the sub-leading distributional identity. Therefore, one should be careful and expand the last term to sub-leading order only. The (sub)$^2$-leading contribution comes from expanding the first term to quadratic order in frequency. Therefore, 
\begin{align}
    \mathcal{R}^{\mu\nu}_{\omega,2} &= \frac{1}{m_1 m_2}\int \hat{d}^4 q e^{-i q\cdot b/\hbar}\hat{\delta}(u_2 \cdot q) \Big\{\hat{\delta}(u_1 \cdot q) \frac{(i k\cdot b/\hbar)^2}{2} - (u_1 \cdot k) (i k \cdot b/\hbar) \hat{\delta}^\prime (u_1 \cdot q) \cr
    &\hspace{7cm}+ \frac{1}{2} (u_1 \cdot k)^2 \hat{\delta}^{\prime\prime}(u_1 \cdot q)\Big\} S^{(0),\mu\nu} \mathcal{A}_4 \,.
\end{align}
We have, for particle 1
\begin{align}
    S^{(0),\mu\nu} &= \frac{1}{p_1 \cdot k} p_1^{(\mu} p_1^{\nu)} - \frac{1}{\tilde{p}_1 \cdot k} \tilde{p}_1^{(\mu} \tilde{p}_1^{\nu)} = -\frac{ \bq^{(\mu}p_1^{\nu)}}{p_1 \cdot \bar{k}} + \frac{(\bq \cdot \bar{k})p_1^{(\mu} p_1^{\nu)}}{(p_1 \cdot \bar{k})^2} \,.
\end{align}
We have the following integrals
\begin{align}
    \mathcal{I}_1^{\mu\nu} &= -\kappa^3(2(p_1 \cdot p_2)^2 - m_1^2 m_2^2)\frac{(\bar{k}\cdot b)^2}{4}  \int \frac{\hat{d}^4 \bq}{\bq^2} e^{-i \bq\cdot b}\hat{\delta}(u_1 \cdot \bq)\hat{\delta}(u_2 \cdot \bq) \Big(-\frac{\bq^{(\mu}p_1^{\nu)}}{p_1 \cdot \bar{k}} + \frac{(\bq \cdot \bar{k})p_1^{(\mu} p_1^{\nu)}}{(p_1 \cdot \bar{k})^2} \Big) \cr
    &= i\kappa^3(2(p_1 \cdot p_2)^2 - m_1^2 m_2^2)\frac{(\bar{k}\cdot b)^2}{8\pi \gamma\beta} \Big(\frac{ b^{(\mu}p_1^{\nu)}}{p_1 \cdot \bar{k}} - \frac{(b \cdot \bar{k})p_1^{(\mu} p_1^{\nu)}}{(p_1 \cdot \bar{k})^2} \Big) \,,
\end{align}
where we have used the integral result of equation \eqref{i2}. Next, we have
\begin{align}
    \mathcal{I}_2^{\mu\nu} &= \kappa^3(2(p_1 \cdot p_2)^2 - m_1^2 m_2^2)\frac{(u_1 \cdot \bar{k})(i \bar{k} \cdot b)}{2}  \int \frac{\hat{d}^4 \bq}{\bq^2} e^{-i \bq\cdot b}\hat{\delta}^\prime(u_1 \cdot \bq)\hat{\delta}(u_2 \cdot \bq) \Big(\frac{\bq^{(\mu}p_1^{\nu)}}{p_1 \cdot \bar{k}} - \frac{(\bq \cdot \bar{k})p_1^{(\mu} p_1^{\nu)}}{(p_1 \cdot \bar{k})^2} \Big) \cr
    &= -\frac{\kappa^3}{4\pi\gamma^3 \beta^3}(2(p_1 \cdot p_2)^2 - m_1^2 m_2^2)\omega b \log{(\omega b)} \Big((\gamma u_2 -u_1)^{(\mu}u_1^{\nu)} - \frac{((\gamma u_2 -u_1) \cdot \bar{k})u_1^{(\mu} u_1^{\nu)}}{(u_1 \cdot \bar{k})} \Big)  \,,
\end{align}
where we have used the integral result of equation \eqref{derintegrals}
and
\begin{align}
    \mathcal{I}_3^{\mu\nu} &= -\kappa^3(2(p_1 \cdot p_2)^2 - m_1^2 m_2^2)\frac{(u_1 \cdot \bar{k})^2}{4}  \int \frac{\hat{d}^4 \bq}{\bq^2} e^{-i \bq\cdot b}\hat{\delta}^{\prime\prime}(u_1 \cdot \bq)\hat{\delta}(u_2 \cdot \bq) \Big(\frac{\bq^{(\mu}p_1^{\nu)}}{p_1 \cdot \bar{k}} - \frac{(\bq \cdot \bar{k})p_1^{(\mu} p_1^{\nu)}}{(p_1 \cdot \bar{k})^2} \Big) \cr
    &=-\kappa^3 (2(p_1 \cdot p_2)^2 - m_1^2 m_2^2)\frac{(u_1 \cdot \bar{k})}{2\gamma^2 \beta^2 }  \int \hat{d}^2 \bq_\perp e^{-i \bq_\perp \cdot b} \frac{1}{(q_\perp^2)^2}\Big(\bq_\perp^{(\mu}u_1^{\nu)}- \frac{(\bq_\perp \cdot \bar{k})u_1^{(\mu} u_1^{\nu)}}{(u_1 \cdot \bar{k})} \Big) \cr
    &\rightarrow \mathcal{O}(\omega) \,,
\end{align}
using the integral result of equation \eqref{i3}.
Therefore we collect the log terms and upon simplifying the $\omega \log{\omega}$ terms of radiation kernel w.r.t particle 1 from the quantum soft theorems is given by
\begin{align} 
    \mathcal{R}_{\omega\log{\omega}}^{\mu\nu} &= \frac{\kappa^3 m_1 m_2}{4\pi \gamma^3 \beta^3} (\omega b) \log{ (\omega b)}  \gamma (2\gamma^2 - 3) \Big( u_1^{(\mu} u_2^{\nu)} - \frac{(u_2 \cdot \bar{k})}{(u_1 \cdot \bar{k})} u_1^{(\mu} u_1^{\nu)}  \Big) \,,
\end{align}
which matches with the tree-level contribution to the $\omega \log{\omega}$ term and the log terms of (sub)$^2$-leading order soft expansion of the radiation kernel. The rest of the terms obtained using quantum soft theorems also match with the soft expansion of the radiation kernel in equation \eqref{exactkernel}.

\section{Evaluation of Integrals}\label{allorderintegrals}
In this appendix, we perform the integrals required to calculate the various terms of the soft radiation kernel that appear in the main text. The range of integration is $\omega < |q_\perp| < b^{-1}$ in all the integrals, where $k^\mu = \omega(1,\hat{n})$.\\
We start with the following integral:
\begin{align}
I_1 &=\int \hat{d}^4 \bar{q} e^{-i b\cdot \bar{q}} \hat{\delta} (u_1 \cdot \bar{q}) \hat{\delta}(u_2 \cdot \bar{q}) \frac{1}{\bq^2} \cr
&= -\frac{1}{\gamma\beta}\int \hat{d}^2 \bq_\perp \frac{e^{i b \cdot \bq_\perp}}{\bq_\perp^2} \,.
\end{align}
The two-dimensional integral over $\bq_\perp$ is easily done using polar coordinates. Let the magnitude of $\bq_\perp$ be $r$ and orient
the $x$ and $y$ axes so that $b \cdot \bq_\perp = |b|r \cos{\theta}$. Therefore the integral becomes
\begin{align} \label{i1}
	I_1 &= -\frac{1}{2\pi }\int dr \frac{J_0(b |r| )}{r} = \frac{1}{2\pi} \log{(\omega b)} \,.
\end{align}
Next, we consider
\begin{align} \label{i2}
	I_2^\mu &= \int \hat{d}^4 \bar{q} e^{-i b\cdot \bar{q}} \hat{\delta} (u_1 \cdot \bar{q}) \hat{\delta}(u_2 \cdot \bar{q}) \frac{\bq^\mu}{\bq^2} \cr
&= -\frac{1}{\gamma\beta}\int \hat{d}^2 \bq_\perp e^{i b \cdot \bq_\perp}\frac{\bq_\perp^\mu}{\bq_\perp^2} \cr
&= \frac{i}{\gamma\beta} \frac{\partial}{\partial b_\perp^\mu} \int \hat{d}^2 \bq_\perp \frac{e^{i b \cdot \bq_\perp}}{\bq_\perp^2} = \frac{i}{2\pi \gamma \beta} \frac{b^\mu}{b^2} \,,
\end{align}
where equation \eqref{i1} is used and $\frac{\hat{b}^\mu}{|b|}= -\frac{b^\mu}{b^2}$.\\
We consider the integral
\begin{align}
	I_3^\mu &= \int \hat{d}^4 \bar{q} e^{-i b\cdot \bar{q}} \hat{\delta}^{(n)} (u_1 \cdot \bar{q}) \hat{\delta}(u_2 \cdot \bar{q}) \frac{\bq^\mu}{\bq^2} \,,
\end{align}
where $(n)$ denotes the number of derivatives acting over the on-shell delta function.  To simplify this,  we shall decompose the momentum $\bq ^\mu$ along $u_{1,2}$ and in the transverse direction 
\begin{align}
	\bq^\mu = \alpha_1 u_1^\mu + \alpha_2 u_2^\mu + \bq_\perp^\mu \,, \qquad  u_i \cdot \bq_\perp =0\,,\label{massless_decomp}
\end{align}
where the coefficients are given by
\begin{align}
	\alpha_1 = \frac{1}{\gamma^2 \beta^2} [\gamma x_2 - x_1]\,,\quad  \alpha_2 = \frac{1}{\gamma^2 \beta^2} [\gamma x_1 - x_2] \,,
\end{align}
with $x_{1,2}:=(u_{1,2} \cdot \bq) $.  Due to this change of variables,  the measure transforms as follows
\begin{align}
	\hat{d}^4 \bq =\frac{1}{\gamma \beta} \hat{d}^2 \bq_\perp dx_1 dx_2 \,.
\end{align}
In terms of $x_{1,2}$ and $\bq_\perp$ variables, we rewrite
\begin{align}
	I_3^\mu &= \frac{1}{\gamma\beta}\int \hat{d}^2 \bar{q}_\perp \hat{d}x_1 \hat{d}x_2 e^{i b\cdot \bar{q}_\perp} \hat{\delta}^{(n)} (x_1) \hat{\delta}(x_2) \frac{\bq^\mu}{\bq^2} \,.
\end{align}
Integrating by parts, we have
\begin{align} 
	I_3^\mu &= (-1)^{n}\frac{1}{\gamma\beta}\int \hat{d}^2 \bar{q}_\perp e^{i b\cdot \bar{q}_\perp} \frac{\partial^n}{\partial x_1^n} \left( \frac{\bq^\mu}{\bq^2} \right)\Big|_{x_1=x_2=0} \,.
\end{align}
\begin{equation} \label{derintegrals}
	I_3^\mu =\begin{cases}
    (-1)^{n}\frac{n!}{2\gamma\beta}\int \hat{d}^2 \bar{q}_\perp e^{i b\cdot \bar{q}_\perp} \frac{\bq_\perp^\mu}{\bq_\perp^2}\left[\left(\frac{1}{\sqrt{\bq_\perp^2 \gamma^2\beta^2}}\right)^n + \left(\frac{-1}{\sqrt{\bq_\perp^2 \gamma^2\beta^2}}\right)^n \right], & \text{if $n \geq 2$}.\\
    \frac{1}{2\pi\gamma^3\beta^3} \log{(\omega b)} (\gamma u_2 -u_1)^\mu, & \text{if $n=1$}. \\
    \frac{i}{2\pi \gamma \beta} \frac{b^\mu}{b^2} , & \text{if $n=0$}.
  \end{cases}
\end{equation}
where
\begin{align}
	\frac{\partial^n}{\partial x_1^n}\left( \frac{1}{\bq^2} \right)\Big|_{x_1=x_2=0} &= \frac{n!}{2\bq_\perp^2} \left[\left(\frac{1}{\sqrt{\bq_\perp^2 \gamma^2\beta^2}}\right)^n + \left(\frac{-1}{\sqrt{\bq_\perp^2 \gamma^2\beta^2}}\right)^n \right]
\end{align}
and
\begin{align}
	\frac{\partial}{\partial x_1} \bq^\mu\Big|_{x_1=x_2=0} = \frac{1}{\gamma^2 \beta^2} (\gamma u_2 - u_1)^\mu
\end{align}
Therefore the first integral of equation \eqref{derintegrals} is evaluated as
\begin{align} \label{i3}
	I_{3,1}^\mu &= (-1)^{n}(-i)\frac{n!}{2\gamma\beta} \frac{\partial}{\partial b^\mu}\int \hat{d}^2 \bar{q}_\perp e^{i b\cdot \bar{q}_\perp} \frac{1}{\bq_\perp^2}\left[\left(\frac{1}{\sqrt{\bq_\perp^2 \gamma^2\beta^2}}\right)^n + \left(\frac{-1}{\sqrt{\bq_\perp^2 \gamma^2\beta^2}}\right)^n \right] \cr
	&=(-1)^{n}(-i)\frac{1}{4\pi \gamma\beta} \frac{\partial}{\partial b^\mu} \Bigg[-\omega^{-n} \Gamma (n) \left(\left(\frac{1}{\sqrt{ \gamma^2\beta^2}}\right)^n + \left(\frac{-1}{\sqrt{ \gamma^2\beta^2}}\right)^n\right) \cr
	&\hspace{6cm} \times \left((\omega b)^n \, {}_1 F_2 \left(-\frac{n}{2};1,1-\frac{n}{2};-\frac{1}{4}\right)-1 \right)\Bigg] \cr
	&= (-1)^{n+1} i\frac{n!  \, b^\mu}{4\pi \gamma\beta b^2} \Bigg[b^n  \left(\left(\frac{1}{\sqrt{\gamma^2\beta^2}}\right)^n + \left(\frac{-1}{\sqrt{\gamma^2\beta^2}}\right)^n\right) \cr
	&\hspace{7cm} \times {}_1 F_2 \left(-\frac{n}{2};1,1-\frac{n}{2};-\frac{1}{4}\right)\Bigg] \,,
\end{align}
where ${}_p F_q (a;b;z)$ is the generalized hypergeometric function.\\

Lastly, we evaluate the integral
\begin{align}
	I_4^\mu &= \int \hat{d}^4 \bar{q} e^{-i b\cdot \bar{q}} \hat{\delta}^{(n)} (u_1 \cdot \bar{q}) \hat{\delta}(u_2 \cdot \bar{q}) \frac{\bq^\mu}{(\bq^2)^m} \,,
\end{align}
where $(n)$ denotes the number of derivatives acting over the on-shell delta function and $m \geq 2$.
In terms of $x_{1,2}$ and $\bq_\perp$ variables, we have the following integral
\begin{align}
	I_4^\mu &= \frac{1}{\gamma\beta}\int \hat{d}^2 \bar{q}_\perp \hat{d}x_1 \hat{d}x_2 e^{i b\cdot \bar{q}_\perp} \hat{\delta}^{(n)} (x_1) \hat{\delta}(x_2) \frac{\bq^\mu}{(\bq^2)^m} \,  \,,  \, ~~~~  m \geq 2 \,.
\end{align}
Integrating by parts, we have
\begin{align} 
	I_4^\mu &= (-1)^{n}\frac{1}{\gamma\beta}\int \hat{d}^2 \bar{q}_\perp e^{i b\cdot \bar{q}_\perp} \frac{\partial^n}{\partial x_1^n} \left( \frac{\bq^\mu}{(\bq^2)^m} \right)\Big|_{x_1=x_2=0} \,.
\end{align}
\begin{equation} \label{derintegrals2}
	I_4^\mu =\begin{cases}
     (-1)^{n} \frac{\displaystyle n! \prod_{k=2}^{m} (n+ 2k-2)}{\displaystyle 2 \prod_{k=2}^{m} (2k-2) \gamma\beta}\int \hat{d}^2 \bar{q}_\perp e^{i b\cdot \bar{q}_\perp} \frac{\bq_\perp^\mu}{(\bq_\perp^2)^m}\left[\left(\frac{1}{\sqrt{\bq_\perp^2 \gamma^2\beta^2}}\right)^n + \left(\frac{-1}{\sqrt{\bq_\perp^2 \gamma^2\beta^2}}\right)^n \right], & \text{if $n \geq 2$}.\\
    \frac{b^{2m-2}}{4\pi (m-1)\gamma^3\beta^3} \, {}_1 F_2 \left(1-m;1,2-m;-\frac{1}{4}\right) (\gamma u_2 -u_1)^\mu, & \text{if $n=1$}. \\
    \frac{i b^{2m-4}}{2\pi \gamma \beta}  \, {}_1 F_2 \left(1-m;1,2-m;-\frac{1}{4}\right) b^\mu , & \text{if $n=0$}.
  \end{cases}
\end{equation}
where
\begin{align}
	\frac{\partial^n}{\partial x_1^n}\left( \frac{1}{(\bq^2)^m} \right)\Big|_{x_1=x_2=0} &= \frac{n! \displaystyle \prod_{k=2}^{m} (n+ 2k-2) }{\displaystyle 2 \prod_{k=2}^{m} (2k-2) (\bq_\perp^2)^m} \left[\left(\frac{1}{\sqrt{\bq_\perp^2 \gamma^2\beta^2}}\right)^n + \left(\frac{-1}{\sqrt{\bq_\perp^2 \gamma^2\beta^2}}\right)^n \right] \,.
\end{align}
Therefore the first integral of equation \eqref{derintegrals2} is evaluated as
\begin{align} \label{i4}
	I_{4,1}^\mu &= (-1)^{n}(-i)\frac{n! \displaystyle \prod_{k=2}^{m} (n+ 2k-2) }{\displaystyle 2 \prod_{k=2}^{m} (2k-2) \gamma\beta} \frac{\partial}{\partial b^\mu}\int \hat{d}^2 \bar{q}_\perp e^{i b\cdot \bar{q}_\perp} \frac{1}{(\bq_\perp^2)^m}\left[\left(\frac{1}{\sqrt{\bq_\perp^2 \gamma^2\beta^2}}\right)^n + \left(\frac{-1}{\sqrt{\bq_\perp^2 \gamma^2\beta^2}}\right)^n \right] \cr
	&=(-1)^{n}(-i)\frac{n! \displaystyle \prod_{k=2}^{m} (n+ 2k-2) }{\displaystyle 4\pi (2m+n-2) \prod_{k=2}^{m} (2k-2) \gamma\beta}  \frac{\partial}{\partial b^\mu} \Bigg[-b^{2m+n-2} \left(\left(\frac{1}{\sqrt{ \gamma^2\beta^2}}\right)^n + \left(\frac{-1}{\sqrt{ \gamma^2\beta^2}}\right)^n\right) \cr
	&\hspace{6cm} \times \, {}_1 F_2 \left(-m-\frac{n}{2}+1;1,1-m-\frac{n}{2}+2;-\frac{1}{4}\right)\Bigg] \cr
	&= (-1)^{n+1} i\frac{n! \displaystyle \prod_{k=2}^{m} (n+ 2k-2)  \, b^\mu}{\displaystyle 4\pi \prod_{k=2}^{m} (2k-2) \gamma\beta } \Bigg[b^{2m+n-4} \left(\left(\frac{1}{\sqrt{ \gamma^2\beta^2}}\right)^n + \left(\frac{-1}{\sqrt{ \gamma^2\beta^2}}\right)^n\right) \cr
	&\hspace{6cm} \times \, {}_1 F_2 \left(-m-\frac{n}{2}+1;1,1-m-\frac{n}{2}+2;-\frac{1}{4}\right)\Bigg] \,.
\end{align}
The main text also involves higher-rank integrals of the following form which can be written in terms of derivatives w.r.t $b^\mu$:
\begin{align}
    I_5^{\mu_1 \mu_2 \cdots \mu_r} &= \int \hat{d}^4 \bar{q} e^{-i b\cdot \bar{q}} \hat{\delta}^{(n)} (u_1 \cdot \bar{q}) \hat{\delta}(u_2 \cdot \bar{q}) \frac{\bq^{\mu_1}\bq^{\mu_2}\cdots \bq^{\mu_r}} {(\bq^2)^m} \cr
    &= (-i\partial_b^{\mu_1}) (-i\partial_b^{\mu_2})\cdots (-i\partial_b^{\mu_{r-1}}) \int \hat{d}^4 \bar{q} e^{-i b\cdot \bar{q}} \hat{\delta}^{(n)} (u_1 \cdot \bar{q}) \hat{\delta}(u_2 \cdot \bar{q}) \frac{\bq^{\mu_r}} {(\bq^2)^m}\,.
\end{align}
The results must lie in the plane orthogonal to both $u_1$ and $u_2$. Therefore we use the projected metric \cite{Maybee:2019jus,Vines:2017hyw,Liu:2021zxr}
\begin{align}
    \frac{\partial}{\partial b_\mu} b^\nu &= \Pi^{\mu\nu} = \eta^{\mu\nu} + \frac{1}{\gamma^2 \beta^2} \left(u_1^\mu (u_1 -\gamma u_2)^\nu + u_2^\mu (u_2 -\gamma u_1)^\nu \right) 
\end{align}
to generate the integrals of any rank. For example,
\begin{align}
    I_6^{\mu\nu\rho} &=\int \hat{d}^4 \bar{q} e^{-i b\cdot \bar{q}} \hat{\delta}^{(n)} (u_1 \cdot \bar{q}) \hat{\delta}(u_2 \cdot \bar{q}) \frac{\bq^{\mu}\bq^{\nu}\bq^{\rho}} {(\bq^2)^m} \cr
    &= (-i\partial_b^{\mu}) (-i\partial_b^{\nu}) I_4^{\rho} \,.
\end{align}
Therefore,
\begin{equation} \label{derintegrals6}
	I_6^{\mu\nu\rho} =\begin{cases}
     (-1)^{n+2} i (2m+n-4)\frac{\displaystyle n! \prod_{k=2}^{m} (n+ 2k-2) }{\displaystyle 4\pi \prod_{k=2}^{m} (2k-2) \gamma\beta } \Bigg\{b^{2m+n-8} \left(\left(\frac{1}{\sqrt{ \gamma^2\beta^2}}\right)^n + \left(\frac{-1}{\sqrt{ \gamma^2\beta^2}}\right)^n\right) \\
	\times \, {}_1 F_2 \left(-m-\frac{n}{2}+1;1,1-m-\frac{n}{2}+2;-\frac{1}{4}\right)\Bigg\} \left[(2m+n-6) b^{\mu} b^\nu b^\rho + b^2 b^{(\mu} \Pi^{\nu\rho)} \right], & \text{if $n \geq 2$}.\\
    \\
    \frac{-b^{2m-6}}{2\pi\gamma^3\beta^3} \, {}_1 F_2 \left(1-m;1,2-m;-\frac{1}{4}\right) (\gamma u_2 -u_1)^{(\mu} \left[(2m-4) b^{\nu} b^{\rho)} + b^2 \Pi^{\nu\rho)} \right], & \text{if $n=1$}. \\
    \frac{-i (2m-4) b^{2m-8}}{2\pi \gamma \beta}  \, {}_1 F_2 \left(1-m;1,2-m;-\frac{1}{4}\right) \left[(2m-6) b^{\mu} b^\nu b^\rho + b^2 b^{(\mu} \Pi^{\nu\rho)} \right] , & \text{if $n=0$}.
  \end{cases}
\end{equation}

\section{Detailed derivations of the terms appearing in (sub)$^n$-leading order soft radiation} \label{subnsection_detailed}
In this appendix, we give a detailed derivation of the steps in computing the classical (sub)$^n$-leading order soft radiation from the quantum soft theorems which appeared in Section \ref{subnsection}.
First, let us evaluate the soft operators' action on $\mathcal{A}_4$ that appeared in the main text. We consider the contribution from particle 1 for now. 
The action of the soft operators on the numerator of the amplitudes is given by
\begin{align}
    \kappa \frac{J_1^{\mu\rho}k_\rho J_1^{\nu\sigma} k_\sigma}{p_1 \cdot k}\Big(k \cdot \frac{\partial}{\partial p_i} \Big)^2  [\mathcal{A}_4]_{N} &= \kappa^3  \frac{k_\rho k_\sigma}{2q_2^2 (p_1 \cdot k)} \Big(p_1 \wedge \frac{\partial}{\partial p_1} \Big)^{\mu\rho} \Big(p_1 \wedge \frac{\partial}{\partial p_1} \Big)^{\nu\sigma} \Big(k \cdot \frac{\partial}{\partial p_1} \Big) \cr
    &\hspace{0.5cm}\Big[(k \cdot \tilde{p}_1)(m_2^2 - p_2 \cdot \tilde{p}_2) + (k \cdot \tilde{p}_2)(p_2 \cdot \tilde{p}_1) + (k \cdot p_2)(\tilde{p}_1 \cdot \tilde{p}_2)\Big]\cr
    &= 0\,.
\end{align}
and
\begin{align}
    \kappa \frac{\tilde{J}_1^{\mu\rho}k_\rho \tilde{J}_1^{\nu\sigma} k_\sigma}{\tilde{p}_1 \cdot k} \Big(k \cdot \frac{\partial}{\partial \tilde{p}_1} \Big)^2 [\mathcal{A}_4]_{N} &= \kappa^3  \frac{k_\rho k_\sigma}{2q_2^2 (\tilde{p}_1 \cdot k)} \Big(\tilde{p}_1 \wedge \frac{\partial}{\partial \tilde{p}_1} \Big)^{\mu\rho} \Big(\tilde{p}_1 \wedge \frac{\partial}{\partial \tilde{p}_1} \Big)^{\nu\sigma} \Big(k \cdot \frac{\partial}{\partial \tilde{p}_1} \Big) \cr
    &\hspace{0.5cm}\Big[(k \cdot p_1)(m_2^2 - p_2 \cdot \tilde{p}_2) + (k \cdot \tilde{p}_2)(p_1 \cdot p_2) + (k \cdot p_2)(p_1 \cdot \tilde{p}_2)\Big]\cr
    &= 0 \,.
\end{align}
The classical contribution comes from the action of the soft operators on the denominator of the amplitude and it is given by
\begin{align} \label{subnclassicalamp2}
     & \frac{\kappa}{2} \frac{J_1^{\mu\rho}k_\rho J_1^{\nu\sigma} k_\sigma}{p_1 \cdot k}\Big(k \cdot \frac{\partial}{\partial p_1} \Big)^{n-2}  \mathcal{A}_4 + \frac{\kappa}{2} \frac{\tilde{J}_1^{\mu\rho}k_\rho \tilde{J}_1^{\nu\sigma} k_\sigma}{\tilde{p}_1 \cdot k}\Big(k \cdot \frac{\partial}{\partial p_1} \Big)^{n-2}  \mathcal{A}_4 \cr
     &= (-1)^{n+1} \kappa^3\frac{2^{n-3}}{(\bq^2)^{n+1}} (\bq \cdot \bar{k})^{n-1} \left(\bq^\mu \bq^\nu ((p_1 \cdot p_2)^2 - \frac{1}{2} m_1^2 m_2^2) + \bq^2 \frac{(p_2 \cdot \bar{k})}{(\bq \cdot \bar{k})}\bq^{(\mu} p_1^{\nu)}\right)\,. 
\end{align}
Therefore, the classical contribution to soft radiation from the action of (sub)$^n$-leading soft operator ($S^{(n),\mu\nu}$) on the four-point amplitude alone is given by
\begin{align} 
     \mathcal{R}^{\mu\nu}_{\omega^{(n-1)},A} &= (-1)^{n+1} \frac{ \kappa^3 m_1 m_2}{4} \int \hat{d}^{4} \bq e^{-i \bq \cdot b} \hat{\delta}(u_1 \cdot \bq) \hat{\delta}(u_2 \cdot \bq)  \frac{2^{n-1}}{(\bq^2)^{n+1}} (\bq \cdot \bar{k})^{n-1} \cr &\hspace{5cm} \times \left(\bq^\mu \bq^\nu \Big(\gamma^2 -\frac{1}{2}\Big) + \bq^2 \frac{(u_2 \cdot \bar{k})}{(\bq \cdot \bar{k})}\bq^{(\mu} u_1^{\nu)}\right)  \,.
\end{align}
This term doesn't give any logarithmic contributions using the integral results of Appendix \ref{allorderintegrals}.
Let us evaluate the soft operators' action on the delta function now. Again we restrict to the contribution from particle 1 where we use the distributional identity of equation \eqref{distrib_identity}. 
We have the following terms that contribute to the classical soft radiation:
\begin{align}
    \mathcal{R}^{\mu\nu}_{\omega^{(n-1)},1} &= \frac{1}{m_1 m_2}\int \hat{d}^{4} q_1 \hat{d}^4 q_2 e^{i q_1 \cdot b} \hat{\delta}(u_1 \cdot q_1) \hat{\delta}(u_2 \cdot q_2) \Big\{-(k \cdot \partial)\hat{\delta}^{(4)}(q_1 + q_2) S^{(n-1),\mu\nu}\mathcal{A}_4 \Big\}\cr
    &= \frac{1}{m_1 m_2}\int \hat{d}^{4} q_1 \hat{d}^4 q_2 e^{i q_1 \cdot b} \hat{\delta}(u_1 \cdot q_1) \hat{\delta}(u_2 \cdot q_2) \Big\{\hat{\delta}^{(4)}(q_1 + q_2- k) \cr
    &\hspace{8cm}- \hat{\delta}^{(4)}(q_1 + q_2)  \Big\} S^{(n-1),\mu\nu} \mathcal{A}_4 \,.
\end{align}
Integrating $q_1$  and relabelling $q_2 \rightarrow q$ and keeping only $\mathcal{O}(\omega^{n-1})$ terms, we have, 
\begin{align}
    \mathcal{R}^{\mu\nu}_{\omega^{(n-1)},1} &= \frac{1}{m_1 m_2} \int \hat{d}^4 q e^{-i q\cdot b}\hat{\delta}(u_2 \cdot q) \Big\{(i k\cdot b)\hat{\delta}(u_1 \cdot q) - (u_1 \cdot k) \hat{\delta}^{\prime}(u_1 \cdot q)\Big\} S^{(n-1),\mu\nu} \mathcal{A}_4 \cr
    &=\frac{1}{m_1 m_2} \int \hat{d}^4 \bq e^{-i \bq\cdot b}\hat{\delta}(u_2 \cdot \bq) \Big\{(i \bar{k}\cdot b)\hat{\delta}(u_1 \cdot \bq) - (u_1 \cdot \bar{k}) \hat{\delta}^{\prime}(u_1 \cdot \bq)\Big\} \cr
    &\hspace{1cm}\times (-1)^{n} \kappa^3\frac{2^{n-1}}{(\bq^2)^{n}} (\bq \cdot \bar{k})^{n-2}\left(\bq^\mu \bq^\nu ((p_1 \cdot p_2)^2 - \frac{1}{2} m_1^2 m_2^2) + \bq^2 \frac{(p_2 \cdot \bar{k})}{(\bq \cdot \bar{k})}\bq^{(\mu} p_1^{\nu)}\right) \,. 
\end{align}
as the classical contribution of $S^{(n-1),\mu\nu}$ on the amplitude is given in equation \eqref{subnclassicalamp2}. This doesn't give any logarithmic contributions using the integral results of Appendix \ref{allorderintegrals}. Therefore, we compute the other terms.
\begin{align}
    \mathcal{R}^{\mu\nu}_{\omega^{(n-1)},2} &= \frac{1}{m_1 m_2} \int \hat{d}^{4} q_1 \hat{d}^4 q_2 e^{i q_1 \cdot b} \hat{\delta}(u_1 \cdot q_1) \hat{\delta}(u_2 \cdot q_2) \frac{1}{2}(k \cdot \partial)^2\hat{\delta}^{(4)}(q_1 + q_2) S^{(n-2),\mu\nu} \mathcal{A}_4 \cr
    &= \frac{1}{m_1 m_2} \int \hat{d}^{4} q_1 \hat{d}^4 q_2 e^{i q_1 \cdot b} \hat{\delta}(u_1 \cdot q_1) \hat{\delta}(u_2 \cdot q_2) \Big\{\hat{\delta}^{(4)}(q_1 + q_2- k) - \hat{\delta}^{(4)}(q_1 + q_2) \cr
    &\hspace{7cm}+ (k \cdot \partial)\hat{\delta}^{(4)}(q_1 + q_2)  \Big\} S^{(n-2),\mu\nu} \mathcal{A}_4 \,.
\end{align}
Integrating $q_1$  and relabelling $q_2 \rightarrow q$ and keeping only $\mathcal{O}(\omega^{n-1})$ terms, we have, 
\begin{align}
    \mathcal{R}^{\mu\nu}_{\omega^{(n-1)},2} &= \frac{1}{m_1 m_2} \int \hat{d}^4 \bq e^{-i \bq\cdot b}\hat{\delta}(u_2 \cdot \bq) \Big\{\frac{(i \bar{k}\cdot b)^2}{2}\hat{\delta}(u_1 \cdot \bq)  -(ib \cdot \bar{k})(u_1 \cdot \bar{k}) \hat{\delta}^{\prime}(u_1 \cdot \bq) \cr
    &\hspace{7cm}+ \frac{1}{2} (u_1 \cdot \bar{k})^2 \hat{\delta}^{\prime\prime}(u_1 \cdot \bq)\Big\} S^{(n-2),\mu\nu} \mathcal{A}_4 \,. \cr
\end{align}
This too doesn't give any logarithmic contributions. Similarly, all the other terms till the action of $S^{(2),\mu\nu}$ do not lead to any log terms using the integral results of Appendix \ref{allorderintegrals}. For instance
\begin{align}
    \mathcal{R}^{\mu\nu}_{\omega^{(n-1)},3} &= \frac{1}{m_1 m_2} \int \hat{d}^{4} q_1 \hat{d}^4 q_2 e^{i q_1 \cdot b} \hat{\delta}(u_1 \cdot q_1) \hat{\delta}(u_2 \cdot q_2) \frac{(-1)^{n-2}(k \cdot \partial)^{n-2}}{(n-2)!}\hat{\delta}^{(4)}(q_1 + q_2) S^{(2),\mu\nu} \mathcal{A}_4 \cr
    &= \frac{1}{m_1 m_2}\int \hat{d}^{4} q_1 \hat{d}^4 q_2 e^{i q_1 \cdot b} \hat{\delta}(u_1 \cdot q_1) \hat{\delta}(u_2 \cdot q_2) \Big\{\hat{\delta}^{(4)}(q_1 + q_2- k) - \hat{\delta}^{(4)}(q_1 + q_2) \cr
    &\hspace{2cm}+(k \cdot \partial)\hat{\delta}^{(4)}(q_1 + q_2) + \cdots - \frac{(-1)^{n-3}}{(n-3)!}(k \cdot \partial)^{n-3}\hat{\delta}^{(4)}(q_1 + q_2) \Big\}  S^{(2),\mu\nu} \mathcal{A}_4 \cr
\end{align}
Integrating $q_1$  and relabelling $q_2 \rightarrow q$ and keeping only $\mathcal{O}(\omega^{n-1})$ terms, we have, 
\begin{align}
    \mathcal{R}^{\mu\nu}_{\omega^{(n-1)},3} &= \frac{1}{m_1 m_2}\int \hat{d}^4 q e^{-i q\cdot b}\hat{\delta}(u_2 \cdot q) \Big\{\frac{(i k\cdot b)^{n-2}}{(n-2)!}\hat{\delta}(u_1 \cdot q)  \cr
    &\hspace{3cm}+\sum_{\substack{r,s \geq 1 \\
\ni (r+s) = n-2}}\frac{(-1)^s}{r! s!} (ib \cdot k)^r (u_1 \cdot k)^s \hat{\delta}^{(s)}(u_1 \cdot q) \cr
    &\hspace{3cm} + \frac{(-1)^{n-2}}{(n-2)!} (u_1 \cdot k)^{n-2} \hat{\delta}^{(n-2)}(u_1 \cdot q)\Big\} S^{(2),\mu\nu} \mathcal{A}_4 \cr
    &= -\frac{1}{m_1 m_2}\kappa^3 \int \hat{d}^4 \bq e^{-i \bq\cdot b}\hat{\delta}(u_2 \cdot \bq) \Big\{\frac{(i \bar{k}\cdot b)^{n-2}}{(n-2)!}\hat{\delta}(u_1 \cdot \bq)  \cr
    &\hspace{1cm}+\sum_{\substack{r,s \geq 1 \\
\ni (r+s) = n-2}}\frac{(-1)^s}{r! s!} (ib \cdot \bar{k})^r (u_1 \cdot \bar{k})^s \hat{\delta}^{(s)}(u_1 \cdot \bq) \cr
    &\hspace{1cm} + \frac{(-1)^{n-2}}{(n-2)!} (u_1 \cdot \bar{k})^{n-2} \hat{\delta}^{(n-2)}(u_1 \cdot \bq)\Big\} \frac{1}{(\bq^2)^3} (\bq \cdot \bar{k}) \cr
    &\hspace{4cm} \times \left(\bq^\mu \bq^\nu ((p_1 \cdot p_2)^2 - \frac{1}{2} m_1^2 m_2^2) + \bq^2 \frac{(p_2 \cdot \bar{k})}{(\bq \cdot \bar{k})}\bq^{(\mu} p_1^{\nu)}\right) \,, \cr
\end{align}
as the classical contribution of $S^{(2),\mu\nu}$ on the amplitude is given in equation \eqref{subsubclassicalamp}. Therefore, we compute the remaining two terms which should give logarithmic contributions.
\begin{align}
    \mathcal{R}^{\mu\nu}_{\omega^{(n-1)},4} &= \frac{1}{m_1 m_2}\int \hat{d}^{4} q_1 \hat{d}^4 q_2 e^{i q_1 \cdot b} \hat{\delta}(u_1 \cdot q_1) \hat{\delta}(u_2 \cdot q_2) \frac{(-1)^{n-1}(k \cdot \partial)^{n-1}}{(n-1)!}\hat{\delta}^{(4)}(q_1 + q_2) S^{(1),\mu\nu} \mathcal{A}_4 \cr
    &= \frac{1}{m_1 m_2} \int \hat{d}^{4} q_1 \hat{d}^4 q_2 e^{i q_1 \cdot b} \hat{\delta}(u_1 \cdot q_1) \hat{\delta}(u_2 \cdot q_2) \Big\{\hat{\delta}^{(4)}(q_1 + q_2- k) - \hat{\delta}^{(4)}(q_1 + q_2) \cr
    &\hspace{1cm}+ (k \cdot \partial)\hat{\delta}^{(4)}(q_1 + q_2) + \cdots - \frac{(-1)^{n-2}}{(n-2)!}(k \cdot \partial)^{n-2}\hat{\delta}^{(4)}(q_1 + q_2) \Big\}  S^{(1),\mu\nu} \mathcal{A}_4 \,. \cr
\end{align}
Integrating $q_1$  and relabelling $q_2 \rightarrow q$ and keeping only $\mathcal{O}(\omega^{n-1})$ terms, we have, 
\begin{align}
    \mathcal{R}^{\mu\nu}_{\omega^{(n-1)},4} &= \frac{1}{m_1 m_2} \int \hat{d}^4 q e^{-i q\cdot b}\hat{\delta}(u_2 \cdot q) \Big\{\frac{(i k\cdot b)^{n-1}}{(n-1)!}\hat{\delta}(u_1 \cdot q)  \cr
    &\hspace{3cm}+\sum_{\substack{r,s \geq 1\\
\ni (r+s) = n-1}}\frac{(-1)^s}{r! s!} (ib \cdot k)^r (u_1 \cdot k)^s \hat{\delta}^{(s)}(u_1 \cdot q) \cr
    &\hspace{3cm} + \frac{(-1)^{n-1}}{(n-1)!} (u_1 \cdot k)^{n-1} \hat{\delta}^{(n-1)}(u_1 \cdot q)\Big\} S^{(1),\mu\nu} \mathcal{A}_4 \,.
\end{align}
From equation \eqref{subsoftclasical}, the classical contribution of $S^{(1),\mu\nu}$ on the amplitude is given by
\begin{align}
   S^{(1),\mu\nu} \mathcal{A}_4 &= -\frac{2\kappa^3}{(\bar{k} \cdot p_1)} \Big[p_1^{(\mu} p_2^{\nu)} (\bar{k} \cdot p_1)(p_1 \cdot p_2) - p_1^{(\mu} p_1^{\nu)} (\bar{k} \cdot p_2)(p_1 \cdot p_2) + \frac{1}{2\bq^2} \bq^\mu \bq^\nu (\bar{k} \cdot p_1)\Big] \,.
\end{align}
Therefore, by substituting the above expression we get
\begin{align}
    \mathcal{R}^{\mu\nu}_{\omega^{(n-1)},4} &= \frac{1}{m_1 m_2} \int \hat{d}^4 \bq e^{-i \bq\cdot b}\hat{\delta}(u_2 \cdot \bq) \Big\{\frac{(i \bar{k}\cdot b)^{n-1}}{(n-1)!}\hat{\delta}(u_1 \cdot \bq)  \cr
    &\hspace{2cm}+\sum_{\substack{r,s \geq 1\\
\ni (r+s) = n-1}}\frac{(-1)^s}{r! s!} (ib \cdot \bar{k})^r (u_1 \cdot \bar{k})^s \hat{\delta}^{(s)}(u_1 \cdot \bq) \cr
    &\hspace{2cm} + \frac{(-1)^{n-1}}{(n-1)!} (u_1 \cdot \bar{k})^{n-1} \hat{\delta}^{(n-1)}(u_1 \cdot \bq)\Big\} \cr
    &\hspace{3cm} \times -\frac{2\kappa^3}{(\bar{k} \cdot p_1)\bq^2} p_1^{(\mu} \Big[p_2^{\nu)} (\bar{k} \cdot p_1)(p_1 \cdot p_2) - p_1^{\nu)} (\bar{k} \cdot p_2)(p_1 \cdot p_2) \Big] \cr
    &\hspace{10cm}+ \mathcal{O}(\omega^{n-1})\cr
    &= \frac{i^{n-1} m_1 m_2 \kappa^3 \gamma}{(n-1)!\pi\gamma\beta} (\omega b)^{n-1} \log{(\omega b)} u_1^{(\mu} \Big(u_2^{\nu)}  - u_1^{\nu)} \frac{(\bar{k} \cdot u_2)}{(\bar{k} \cdot u_1)} \Big) + \mathcal{O}(\omega^{n-1}) \,,
\end{align}
where we have used the integral result of equation \eqref{i1}.
The second integral and the third one do not give log terms following the result of equation \eqref{i3}. We are now left with computing one last term.
\begin{align}
    \mathcal{R}^{\mu\nu}_{\omega^{(n-1)},5} &= \frac{1}{m_1 m_2} \int \hat{d}^{4} q_1 \hat{d}^4 q_2 e^{i q_1 \cdot b} \hat{\delta}(u_1 \cdot q_1) \hat{\delta}(u_2 \cdot q_2) \frac{(-1)^{n}(k \cdot \partial)^{n}}{n!}\hat{\delta}^{(4)}(q_1 + q_2) S^{(0),\mu\nu} \mathcal{A}_4 \cr
    &= \frac{1}{m_1 m_2} \int \hat{d}^{4} q_1 \hat{d}^4 q_2 e^{i q_1 \cdot b} \hat{\delta}(u_1 \cdot q_1) \hat{\delta}(u_2 \cdot q_2) \Big\{\hat{\delta}^{(4)}(q_1 + q_2- k) - \hat{\delta}^{(4)}(q_1 + q_2) \cr
    &\hspace{2cm}+ (k \cdot \partial)\hat{\delta}^{(4)}(q_1 + q_2) + \cdots - \frac{(-1)^{n-1}}{(n-1)!}(k \cdot \partial)^{n-1}\hat{\delta}^{(4)}(q_1 + q_2) \Big\}  S^{(0),\mu\nu} \mathcal{A}_4 \cr
\end{align}
Integrating $q_1$  and relabelling $q_2 \rightarrow q$ and keeping only $\mathcal{O}(\omega^{n-1})$ terms, we have, 
\begin{align}
    \mathcal{R}^{\mu\nu}_{\omega^{(n-1)},5} &= \frac{1}{m_1 m_2} \int \hat{d}^4 q e^{-i q\cdot b}\hat{\delta}(u_2 \cdot q) \Big\{\frac{(i k\cdot b)^{n}}{n!}\hat{\delta}(u_1 \cdot q)  \cr
    &+\sum_{\substack{r,s \\
\ni (r+s) = n}}\frac{(-1)^s}{r! s!} (ib \cdot k)^r (u_1 \cdot k)^s \hat{\delta}^{(s)}(u_1 \cdot q) + \frac{(-1)^{n}}{n!} (u_1 \cdot k)^{n} \hat{\delta}^{(n)}(u_1 \cdot q)\Big\} S^{(0),\mu\nu} \mathcal{A}_4 \,. \cr
\end{align}
We have, for particle 1
\begin{align}
    S^{(0),\mu\nu} &= \frac{1}{p_1 \cdot k} p_1^{(\mu} p_1^{\nu)} - \frac{1}{\tilde{p}_1 \cdot k} \tilde{p}_1^{(\mu} \tilde{p}_1^{\nu)} = -\frac{ \bq^{(\mu}p_1^{\nu)}}{p_1 \cdot \bar{k}} + \frac{(\bq \cdot \bar{k})p_1^{(\mu} p_1^{\nu)}}{(p_1 \cdot \bar{k})^2} \,.
\end{align}
\\
We have the following integrals
\begin{align}
    \mathcal{I}_1^{\mu\nu} &= -\kappa^3(2(p_1 \cdot p_2)^2 - m_1^2 m_2^2)\frac{(i\bar{k}\cdot b)^n}{2n!}  \int \frac{\hat{d}^4 \bq}{\bq^2} e^{-i \bq\cdot b}\hat{\delta}(u_1 \cdot \bq)\hat{\delta}(u_2 \cdot \bq) \Big(\frac{\bq^{(\mu}p_1^{\nu)}}{p_1 \cdot \bar{k}} - \frac{(\bq \cdot \bar{k})p_1^{(\mu} p_1^{\nu)}}{(p_1 \cdot \bar{k})^2} \Big) \cr
    &= -i\kappa^3(2(p_1 \cdot p_2)^2 - m_1^2 m_2^2)\frac{(i \bar{k}\cdot b)^n}{4n!\pi \gamma\beta} \Big(\frac{b^{(\mu}p_1^{\nu)}}{p_1 \cdot \bar{k}} - \frac{(b \cdot \bar{k})p_1^{(\mu} p_1^{\nu)}}{(p_1 \cdot \bar{k})^2} \Big) \,,
\end{align}
using the integral result of equation \eqref{i2}.
Next, we have the integral
\begin{align}
    \mathcal{I}_2^{\mu\nu} &= -\kappa^3(2(p_1 \cdot p_2)^2 - m_1^2 m_2^2)\sum_{\substack{r,s \\
\ni (r+s) = n}}\frac{(-1)^s}{2r! s!} (ib \cdot \bar{k})^r (u_1 \cdot \bar{k})^s \int \frac{\hat{d}^4 \bq}{\bq^2} e^{-i \bq\cdot b}\hat{\delta}^{(s)}(u_1 \cdot \bq)\hat{\delta}(u_2 \cdot \bq) \cr
&\hspace{7cm} \times \Big(\frac{\bq^{(\mu}p_1^{\nu)}}{p_1 \cdot \bar{k}} - \frac{(\bq \cdot \bar{k})p_1^{(\mu} p_1^{\nu)}}{(p_1 \cdot \bar{k})^2} \Big) \cr
    &= \kappa^3 (2(p_1 \cdot p_2)^2 - m_1^2 m_2^2)\sum_{\substack{r,s \\
\ni (r+s) = n}}\frac{(-1)^s}{4r! \gamma \beta} (ib \cdot \bar{k})^r  (u_1 \cdot \bar{k})^{s-1} \cr
&\hspace{1cm}\int \hat{d}^2 \bq_\perp e^{-i \bq_\perp \cdot b} \frac{1}{\bq_\perp^2} \Big[\frac{1}{(\bq_\perp^2 \gamma^2\beta^2)^{s/2}} + \Big(\frac{-1}{\sqrt{\bq_\perp^2 \gamma^2\beta^2}}\Big)^s \Big] \Big(\bq_\perp^{(\mu}u_1^{\nu)}- \frac{(\bq_\perp \cdot \bar{k})u_1^{(\mu} u_1^{\nu)}}{(u_1 \cdot \bar{k})} \Big) \cr
    &-\kappa^3(2(p_1 \cdot p_2)^2 - m_1^2 m_2^2)\frac{1}{2(n-1)! \gamma \beta} (ib \cdot \bar{k})^{n-1} \int \hat{d}^2 \bq_\perp e^{-i \bq_\perp \cdot b} \frac{1}{\bq_\perp^2} \cr
    &\hspace{5cm}\frac{\partial}{\partial (u_1 \cdot \bq)}\Big(\bq^{(\mu}u_1^{\nu)} - \frac{(\bq \cdot \bar{k})u_1^{(\mu} u_1^{\nu)}}{(u_1 \cdot \bar{k})} \Big) \cr
    &= - \kappa^3(2(p_1 \cdot p_2)^2 - m_1^2 m_2^2)\frac{i^{n-1}}{4\pi(n-1)!\gamma^3 \beta^3} (\omega b)^{n-1}\log{(\omega b)} \cr
    &\hspace{5cm} \times \Big((\gamma u_2 - u_1)^{(\mu}u_1^{\nu)}- \frac{((\gamma u_2 -u_1) \cdot \bar{k})u_1^{(\mu} u_1^{\nu)}}{(u_1 \cdot \bar{k})} \Big) + \mathcal{O}(\omega^{n-1})\,,
\end{align}
where we have used the integral result of equation \eqref{derintegrals},
and lastly
\begin{align}
    \mathcal{I}_3^{\mu\nu} &= -\kappa^3(2(p_1 \cdot p_2)^2 - m_1^2 m_2^2)\frac{(-1)^n(u_1 \cdot \bar{k})^n}{2n!}  \int \frac{\hat{d}^4 \bq}{\bq^2} e^{-i \bq\cdot b}\hat{\delta}^{(n)}(u_1 \cdot \bq)\hat{\delta}(u_2 \cdot \bq) \cr
    &\hspace{7cm} \Big(\frac{\bq^{(\mu}p_1^{\nu)}}{p_1 \cdot \bar{k}} - \frac{(\bq \cdot \bar{k})p_1^{(\mu} p_1^{\nu)}}{(p_1 \cdot \bar{k})^2} \Big) \cr
    &= \kappa^3(2(p_1 \cdot p_2)^2 - m_1^2 m_2^2)\frac{(-1)^n(u_1 \cdot \bar{k})^n}{2n! \gamma\beta}  \int \hat{d}^2 \bq_\perp e^{-i \bq_\perp \cdot b} \cr
    &\hspace{5cm} \frac{\partial^n}{\partial (u_1 \cdot \bq)^n}\frac{1}{\bq^2}\Big(\frac{\bq^{(\mu}p_1^{\nu)}}{p_1 \cdot \bar{k}} - \frac{(\bq \cdot \bar{k})p_1^{(\mu} p_1^{\nu)}}{(p_1 \cdot \bar{k})^2} \Big) \,.
\end{align}
Using the integral result of equation \eqref{derintegrals}, we get
\begin{equation}
	\mathcal{I}_3^{\mu\nu} =\begin{cases}
    \mathcal{O}(\omega^{n-1}), & \text{if $n \geq 2$}.\\
    - \kappa^3(2(p_1 \cdot p_2)^2 - m_1^2 m_2^2)\frac{1}{4\pi\gamma^3 \beta^3} \log{(\omega b)} \Big((\gamma u_2 - u_1)^{(\mu}u_1^{\nu)}- \frac{((\gamma u_2 -u_1) \cdot \bar{k})u_1^{(\mu} u_1^{\nu)}}{(u_1 \cdot \bar{k})} \Big) , & \text{if $n=1$}.
  \end{cases}
\end{equation}
Here the $\log(\omega b)$ contribution comes only from $n=1$.\\
Therefore we collect the log terms and upon simplifying the $\omega^{n-1} \log{\omega}$ terms of radiation kernel w.r.t particle 1 from the quantum soft theorems is given by
\begin{align}
    \mathcal{R}^{\mu\nu}_{\omega^{n-1}\log{\omega}} &= \frac{i^{n-1}m_1 m_2\kappa^3}{4\pi(n-1)!\gamma^3\beta^3}\gamma (2\gamma^2 - 3) (\omega b)^{n-1} \log{(\omega b)} \Big( u_1^{(\mu} u_2^{\nu)} - \frac{(u_2 \cdot \bar{k})}{(u_1 \cdot \bar{k})} u_1^{(\mu} u_1^{\nu)}  \Big) \,.
\end{align}

\end{appendix}
\bibliographystyle{JHEP}

\bibliography{subnsoft}

\end{document}